\documentclass[12pt]{iopart}
\usepackage{pdflscape}
\usepackage{epsfig}
\usepackage{iopams}

\renewcommand{\theequation}{\thesection.\arabic{equation}}
\newcommand{\be}{\begin{equation}}
\newcommand{\ee}{\end{equation}}

\newcommand{\ba}{\begin{array}}
\newcommand{\ea}{\end{array}}
\newcommand{\bea}{\begin{eqnarray}}
\newcommand{\eea}{\end{eqnarray}}

\newcommand{\sdir}{\ensuremath{\rlap{\raisebox{.15ex}{$\mskip
6.5mu\scriptstyle+ $}}\supset}}

\newtheorem{proposition}{Proposition}

\begin{document}

\title[Supersymmetric version of the equations of conformally parametrized surfaces]{Supersymmetric version of the equations of conformally parametrized surfaces}
\author{S Bertrand$^1$, A M Grundland$^{2,3}$ and A J Hariton$^2$}

\address{$^1$ Department of Mathematics and Statistics, Universit\'e de Montr\'eal,\\ Montr\'eal CP 6128 Succ. Centre-Ville (QC) H3C 3J7, Canada}
\address{$^2$ Centre de Recherches Math\'ematiques, Universit\'e de Montr\'eal,\\ Montr\'eal CP 6128 Succ. Centre-Ville (QC) H3C 3J7, Canada}
\address{$^3$ Department of Mathematics and Computer Science, Universit\'e du Qu\'ebec, Trois-Rivi\`eres, CP 500 (QC) G9A 5H7, Canada}
\ead{bertrans@crm.umontreal.ca, grundlan@crm.umontreal.ca, hariton@crm.umontreal.ca}

\begin{abstract} In this paper, we formulate a supersymmetric extension of the Gauss-Weingarten and Gauss-Codazzi equations for conformally parametrized surfaces immersed in a Grassmann superspace. We perform this analysis using a superspace-superfield formalism together with a supersymmetric version of a moving frame on a surface. In constrast to the classical case, where we have three Gauss-Codazzi equations, we obtain six such equations in the supersymmetric case. We determine the Lie symmetry algebra of the classical Gauss-Codazzi equations to be infinite-dimensional and perform a subalgebra classification of the one-dimensional subalgebras of its largest finite-dimensional subalgebra. We then compute a superalgebra of Lie point symmetries of the supersymmetric Gauss-Codazzi equations and classify the one-dimensional subalgebras of this superalgebra into conjugacy classes. We then use the symmetry reduction method to find invariants, orbits and reduced systems for two one-dimensional subalgebras in the classical case and three one-dimensional subalgebras in the supersymmetric case. Through the solutions of these reduced systems, we obtain explicit solutions and surfaces of the classical and supersymmetric Gauss-Codazzi equations. We provide a geometrical interpretation of the results. 

\paragraph{}Keywords: conformally parametrized surfaces, supersymmetric models, Lie superalgebras, symmetry reduction. 
\end{abstract}
\pacs{02.20.Sv, 02.40.Hw, 11.15.Wx}
\ams{35Q53, 53A05, 22E70}

\maketitle

\section{Introduction}
The theory of supersymmetry has been studied extensively over the past decades and has generated a great deal of interest and activity in several areas of mathematics and physics. Applications have been numerous in the field of particle physics (see e.g. \cite{Binetruy}-\cite{Crombrugghe}) but it has also been applied to classical fluid dynamics (\cite{Jackiw}-\cite{Fatyga}). Various approaches have been used to construct supersoliton solutions, such as the inverse scattering method, B\"acklund and Darboux transformations for odd and even superfields, Lax formalism in a superspace and generalized versions of the symmetry reduction method. A number of soliton and super multi-soliton solutions were determined by a Crum-type transformation \cite{Matveev}-\cite{Siddiq05} and it was found that there exist infinitely many local conserved quantities. A connection was established between the super-Darboux transformations and super-B\"acklund transformations which allow for the construction of supersoliton solutions \cite{Siddiq06}-\cite{Tian}.

\paragraph{}In differential geometry, parametrized surfaces are described in terms of moving frames satisfying the Gauss-Weingarten (GW) equations, which are linear differential equations. Their compatibility conditions are the Gauss-Codazzi (GC) equations. A representation of nonlinear equations in the form of the GC equations is the starting point in the theory of integrable (soliton) surfaces arising from infinitesimal deformations of integrable differential equations and describing the behaviour of soliton solutions. The construction and analysis of such surfaces associated with integrable systems in several areas of mathematical physics provides new tools for the investigation of nonlinear phenomena described by these systems. In this setting, it is our objective to perform a systematic analysis of a supersymmetric (SUSY) version of the GC equations. The formulation of a SUSY extension of the GW and GC equations has already been accomplished for the specific case of bosonic Grassmann sigma models \cite{Sasaki}-\cite{Witten}. It would be of considerable interest to consider such an extension for general case of the GW and GC equations.

\paragraph{}The purpose of this paper is to formulate a SUSY extension of the GW and GC equations for the case of conformally parametrized surfaces in $\mathbb{R}^3$. The SUSY version of these equations is formulated through the use of a superspace-superfield formalism. The considered surfaces are parametrized by the vector field $\mathcal{F}$ and the normal vector field $\mathcal{N}$, which are replaced in the SUSY version by their corresponding bosonic vector superfields $F$ and $N$ in a superspace $\mathbb{R}^{(1,1\vert2)}$. This allows us to formulate the SUSY extension of the structural equations for the immersion of conformally parametrized surfaces explicitly in matrix form. We establish an explicit form of the SUSY GW equations satisfied by the moving frame on these surfaces. The result is independent of the parametrization. This allows us to examine various geometric properties of the studied immersions, such as the first and second fundamental forms of the surfaces (and therefore the mean and Gaussian curvatures).

\paragraph{}Once we have established the SUSY extension of the GW and GC equations, we compute a Lie symmetry superalgebra and classify its one-dimensional subalgebras into conjugacy classes. We then use a generalized version of the symmetry reduction method to determine invariant solutions of our SUSY model. Some geometrical aspects of the obtained results are explored. We demonstrate that the SUSY GW equations for the superframe on the surface resemble the linear system which appears in the soliton theory of the immersion of surfaces in Lie algebras. 

\paragraph{}The paper is organized as follows. The symmetry algebra of the classical GC equations is determined and a subalgebra classification of its one-dimensional subalgebras is performed in section 2. In section 3, we recall the basic properties of Grassmann algebras and Grassmann variables and introduce the notation that will be used in what follows. In section 4, we construct the supersymmetric extensions of the Gauss-Weingarten and Gauss-Codazzi equations. In section 5, we discuss certain geometric aspects of the conformally parametrized supersymmetric surface. We provide expressions for the first and second fundamental forms and the Gaussian and mean curvatures, which are required for a geometrical interpretation of the invariant solutions. In section 6, we determine a Lie superalgebra of symmetries of the supersymmetric Gauss-Codazzi equations. Section 7 involves a classification of the one-dimensional subalgebras of the Lie superalgebra into conjugacy classes. In section 8, we provide examples of invariant solutions of the supersymmetric Gauss-Codazzi equations obtained by the symmetry reduction method. Finally, in section 9, we present the conclusions and discuss possible future developments in this field.

\section{Symmetries of conformally parametrized surfaces}
The system of partial differential equations (PDEs) describing the moving frame $\Omega=(\partial \mathcal{F},\bar{\partial}\mathcal{F},\mathcal{N})^T$ on a conformally parametrized surface in $3$-dimensional Euclidean space satisfies the following GW equations
\begin{equation}
\partial\Omega=V_1\Omega,\qquad\bar{\partial}\Omega=V_2\Omega,\label{1.1}
\end{equation}
where the matrices $V_1$ and $V_2$ are given by
\begin{equation}
\hspace{-1.5cm}V_1=\left(\begin{array}{ccc}
\partial u&0&Q\\
0&0&\frac{1}{2}He^u\\
-H&-2Qe^{-u}&0
\end{array}\right),\qquad V_2=\left(\begin{array}{ccc}
0&0&\frac{1}{2}He^u\\
0&\bar{\partial}u&\bar{Q}\\
-2\bar{Q}e^{-u}&-H&0
\end{array}\right).\label{claV1V2}
\end{equation}
Here $\partial$ and $\bar{\partial}$ are the partial derivatives with respect to the complex coordinates $z=x+iy$ and $\bar{z}=x-iy$, respectively. The conformal parametrization of a surface is given by a vector-valued function $\mathcal{F}=(\mathcal{F}_1,\mathcal{F}_2,\mathcal{F}_3)^T:\mathcal{R}\rightarrow\mathbb{R}^3$ (where $\mathcal{R}$ is a Riemann surface) which satisfies the following normalization for the tangent vectors $\partial \mathcal{F}$ and $\bar{\partial}\mathcal{F}$ and the unit normal $N$
\begin{equation}
\begin{array}{lll}
\langle\partial \mathcal{F},\partial \mathcal{F}\rangle=\langle\bar{\partial}\mathcal{F},\bar{\partial}\mathcal{F}\rangle=0,& &\langle\partial \mathcal{F},\bar{\partial}\mathcal{F}\rangle=\frac{1}{2}e^u,\\
\langle\partial \mathcal{F},\mathcal{N}\rangle=\langle\bar{\partial}\mathcal{F},\mathcal{N}\rangle=0,& &\langle \mathcal{N},\mathcal{N}\rangle=1,
\end{array}
\end{equation}
and we define the quantities
\begin{equation}
Q=\langle\partial^2\mathcal{F},\mathcal{N}\rangle\in\mathbb{C},\qquad H=2e^{-u}\langle\partial\bar{\partial}\mathcal{F},\mathcal{N}\rangle\in\mathbb{R}.
\end{equation}
Here the bracket $\langle\cdot,\cdot\rangle$ denotes the scalar product in 3-dimensional Euclidean space $\mathbb{R}^3$
\begin{equation}
\langle a,b\rangle=a_1b_1+a_2b_2+a_3b_3.\label{innerproduct}
\end{equation}
So, the GW equations for a moving frame $\Omega$ on a surface have to obey the following system of equations
\begin{equation}
\hspace{-2cm}\begin{array}{l}
\partial^2\mathcal{F}=\partial u\partial \mathcal{F}+Q\mathcal{N},\qquad \partial\bar{\partial}\mathcal{F}=\frac{1}{2}He^u\mathcal{N},\qquad\bar{\partial}^2\mathcal{F}=\bar{\partial}u\bar{\partial}\mathcal{F}+\bar{Q}\mathcal{N},\\
\partial \mathcal{N}=-H\partial \mathcal{F}-2e^{-u}Q\bar{\partial}\mathcal{F},\qquad \bar{\partial}\mathcal{N}=-2e^{-u}\bar{Q}\partial \mathcal{F}-H\bar{\partial}\mathcal{F}.
\end{array}
\end{equation}
The first and second fundamental forms are given by
\begin{equation}
\hspace{-2.5cm}I=\langle d\mathcal{F},d\mathcal{F}\rangle=\left\langle\frac{e^u}{2}\left(\begin{array}{cc}
0&1\\
1&0
\end{array}\right)\left(\begin{array}{c}
dz\\
d\bar{z}
\end{array}\right),\left(\begin{array}{c}
dz\\
d\bar{z}
\end{array}\right)\right\rangle=e^u\left\langle\left(\begin{array}{c}
dx\\
dy
\end{array}\right),\left(\begin{array}{c}
dx\\
dy
\end{array}\right)\right\rangle,\label{Icla}
\end{equation}
and
\begin{equation}
\hspace{-2.5cm}I\hspace{-0.1cm}I=\langle d^2\mathcal{F},\mathcal{N}\rangle=\left\langle\left(\begin{array}{cc}
Q+\bar{Q}+e^uH&i(Q-\bar{Q})\\
i(Q-\bar{Q})&-(Q+\bar{Q})+e^uH
\end{array}\right)\left(\begin{array}{c}
dx\\
dy
\end{array}\right),\left(\begin{array}{c}
dx\\
dy
\end{array}\right)\right\rangle,\label{IIcla}
\end{equation}
respectively. The principal curvatures $k_1$ and $k_2$ are the eigenvalues of the matrix
\begin{equation}
B=e^{-u}\left(\begin{array}{cc}
Q+\bar{Q}+e^uH&i(Q-\bar{Q})\\
i(Q-\bar{Q})&-(Q+\bar{Q})+e^uH
\end{array}\right).
\end{equation}
We obtain the following expressions for the mean and Gaussian curvatures
\begin{equation}
H=\frac{1}{2}(k_1+k_2)=\frac{1}{2}\mbox{tr}(B),
\end{equation}
\begin{equation}
\mathcal{K}=k_1k_2=\det(B)=H^2-4Q\bar{Q}e^{-2u}.
\end{equation}
Umbilic points on a surface take place when $H^2-\mathcal{K}=0$ which implies that $\vert Q\vert^2=0$. The compatibility conditions of the GW equations (\ref{1.1}) are the GC equations
\begin{equation}
\bar{\partial}V_1-\partial V_2+[V_1,V_2]=0,
\end{equation}
(the bracket $[\cdot,\cdot]$ denotes the commutator) which reduce to the following three differential equations for the quantities $Q$, $H$ and $e^u$
\begin{equation}
\hspace{-1cm}\begin{array}{cr}
\partial\bar{\partial}u+\frac{1}{2}H^2e^u-2Q\bar{Q}e^{-u}=0,&\mbox{(the Gauss equation)}\\
\partial \bar{Q}-\frac{1}{2}e^u\bar{\partial}H=0,\qquad\bar{\partial}Q-\frac{1}{2}e^u\partial H=0.&\mbox{(the Codazzi equations)}
\end{array}
\label{cCeq}
\end{equation}
These equations are the necessary and sufficient conditions for the existence of conformally parametrized surfaces in 3-dimensional Euclidean space $\mathbb{R}^3$ with the fundamental forms given by (\ref{Icla}) and (\ref{IIcla}). A review of systematic computational methods for constructing surfaces for a given moving frame can be found in several books (e.g. \cite{Baer}-\cite{Thorpe}). Equations (\ref{1.1}), (\ref{claV1V2}) and (\ref{cCeq}) allow us to formulate explicitly the structural equations for the immersion directly in matrix terms. However, it is nontrivial to identify those surfaces which have an invariant geometrical characterization \cite{do Carmo},\cite{Bob}. The task of finding an increasing number of solutions of the GW and GC equations is related to the group properties of these systems of equations. Their main advantages appear when group analysis makes it possible to construct regular algorithms for finding certain classes of solutions (describing diverse types of surfaces) without referring to any additional considerations, but proceeding only from the given system of equations. A broad review of recent developments in this subject can be found in several books (see e.g. P. Olver\cite{Olver}, D. Sattinger and O. Weaver \cite{SW}, and G. Bluman and S. Kumai \cite{BK}). The methodological approach adopted here is based on the symmetry reduction method for PDEs invariant under a Lie group $G$ of point transformations.
Using the Maple program, we find that the symmetry group of the classical GC equations (\ref{cCeq}) consists of conformal scaling transformations. The corresponding symmetry algebra $\mathcal{L}$ is spanned by the vector fields
\be
\ba{l}
X(\eta)=\eta(z)\partial_z+\eta'(z)(-2Q\partial_Q-U\partial_U),\\
Y(\zeta)=\zeta(\bar{z})\partial_{\bar{z}}+\zeta'(\bar{z})(-2\bar{Q}\partial_{\bar{Q}}-U\partial_U),\\
e_0=-H\partial_H+Q\partial_Q+\bar{Q}\partial_{\bar{Q}}+2U\partial_U,
\ea\label{1.14}
\ee
where $\eta'(\cdot)$ and $\zeta'(\cdot)$ are the derivatives of $\eta(\cdot)$ and $\zeta(\cdot)$ with respect to their arguments respectively and where we have used the notation $e^u=U$. The commutation relations are
\begin{equation}
\hspace{-2.5cm}\begin{array}{l}
[X(\eta_1),X(\eta_2)]=(\eta'_1\eta_2-\eta_1\eta'_2)\partial_z+(\eta_1''\eta_2-\eta_1\eta_2'')(2Q\partial_Q+U\partial_U),\\
\left[Y(\zeta_1),Y(\zeta_2)\right]=(\zeta_1'\zeta_2-\zeta_1\zeta_2')\partial_{\bar{z}}+(\zeta_1''\zeta_2-\zeta_1\zeta_2'')(2\bar{Q}\partial_{\bar{Q}}+U\partial_U),\\
\left[ X(\eta),Y(\zeta) \right]=0,\qquad [X(\eta),e_0]=0, \qquad [Y(\zeta),e_0]=0.
\end{array}
\end{equation}
Since the vector fields $X(\eta), Y(\zeta)$ and $e_0$ form an Abelian algebra, they determine that the algebra $\mathcal{L}$ can be decomposed as a direct sum of two infinite-dimensional Lie algebras together with a one-dimensional algebra generated by $e_0$, i.e.
\be
\mathcal{L}=\lbrace X(\eta)\rbrace\oplus\lbrace Y(\zeta)\rbrace\oplus\lbrace e_0\rbrace.
\ee
This algebra represents a direct sum of two copies of the Virasoro algebra together with the one-dimensional subalgebra $\lbrace e_0\rbrace$. Assuming that the functions $\eta$ and $\zeta$ are analytic in some open subset $\mathcal{D}\subset\mathbb{C}$, we can develop $\eta$ and $\zeta$ as power series with respect to their arguments and provide a basis for $\mathcal{L}$. The largest finite-dimensional subalgebra $L$ of the algebra $\mathcal{L}$ is spanned by seven generators
\be
\hspace{-1.5cm}\ba{l}
e_0=-H\partial_H+Q\partial_Q+\bar{Q}\partial_{\bar{Q}}+2U\partial_U,\\
e_1=\partial_z,\qquad e_3=z\partial_z-2Q\partial_Q-U\partial_U,\qquad e_5=z^2\partial_z-4zQ\partial_Q-2zU\partial_U,\\
e_2=\partial_{\bar{z}},\qquad e_4=\bar{z}\partial_{\bar{z}}-2\bar{Q}\partial_{\bar{Q}}-U\partial_U,\qquad e_6=\bar{z}^2\partial_{\bar{z}}-4\bar{z}\bar{Q}\partial_{\bar{Q}}-2\bar{z}U\partial_U,
\ea\label{ei}
\ee
with nonzero commutation relations
\be
\ba{lll}
\phantom{~}[e_1,e_3]=e_1,& [e_1,e_5]=-2e_3,& [e_3,e_5]=e_5,\\
\phantom{~}[e_2,e_4]=e_2,& [e_2,e_6]=-2e_4,& [e_4,e_6]=e_6.
\ea
\ee
This 7-dimensional Lie subalgebra $L$ can be decomposed as a direct sum of two simple subalgebras together with a one-dimensional algebra generated by $e_0$,
\be
L=\lbrace e_1, e_3, e_5\rbrace\oplus\lbrace e_2,e_4, e_6\rbrace\oplus\lbrace e_0\rbrace.\label{1.19}
\ee
Therefore, the classification of the subalgebras of $L$ consists of two copies of a 3-dimensional Lie algebra together with the center $\lbrace e_0\rbrace$. This 3-dimensional Lie algebra corresponds to the algebra $A_{3,8}$ in the classification of J. Patera and P. Winternitz \cite{Patera} which is isomorphic to $\mathfrak{su}(1,1)$. The resulting classification of the subalgebras of $L$ into conjugacy classes, performed according to the methods described in \cite{Winternitz}, is given by the following list of representative subalgebras $L_{1,j}$
\be
\hspace{-2.7cm}\ba{lll}
L_{1,0}=\lbrace e_0\rbrace, & \hspace{-1cm}L_{1,1}=\lbrace e_1\rbrace, & \hspace{-2cm}L_{1,2}=\lbrace e_3\rbrace, \qquad L_{1,3}=\lbrace e_1+e_5\rbrace,\hspace{-2cm} \\
L_{1,4}=\lbrace e_2\rbrace, & L_{1,5}=\lbrace e_4\rbrace, & L_{1,6}=\lbrace e_2+e_6\rbrace,  \\
L_{1,7}=\lbrace e_1+\epsilon e_2\rbrace, & L_{1,8}=\lbrace e_1+\epsilon e_4\rbrace, & L_{1,9}=\lbrace e_2+e_6+\epsilon e_1\rbrace, \\
L_{1,10}=\lbrace e_3+\epsilon e_2\rbrace, & L_{1,11}=\lbrace e_3+ae_4\rbrace, & L_{1,12}=\lbrace e_2+e_6+ae_3\rbrace,  \\
L_{1,13}=\lbrace e_1+e_5+\epsilon e_2\rbrace, & L_{1,14}=\lbrace e_1+e_5+ae_4\rbrace, & L_{1,15}=\lbrace e_1+e_5+a(e_2+e_6)\rbrace,  \hspace{-2cm}
\ea\label{1.20}
\ee
where $\epsilon=\pm1$ and $a\neq0$ are parameters. The center of $L$, $\lbrace e_0\rbrace$, can be added to any of the subalgebras given above, say $L_{1,j}=\lbrace A\rbrace$, to produce a twisted subalgebra of the form $L_{1,j}'=\lbrace A+be_0\rbrace$, where $b\neq0$. The symmetry reductions associated with the subalgebras (\ref{1.20}) lead to systems of ordinary differential equations (ODEs). These reduced systems were analyzed systematically as a single generic symmetry reduction in \cite{ConteGrundland}, where the GC equations (\ref{cCeq}) were reduced to the most general Painlev\'e P6 form (containing two or three arbitrary parameters).

\section{Preliminaries on Grassmann algebras}\label{intro}\setcounter{equation}{0}
The mathematical background formalism is based on the theory of supermanifolds as presented in \cite{Cornwell}-\cite{Varadarajan} and can be summarized as follows. The starting point in our consideration is a complex Grassmann algebra $\Lambda$ involving a finite or infinite number of Grassmann generators $(\xi_1,\xi_2,...)$. The number of Grassmann generators of $\Lambda$ is not essential provided that there is a sufficient number of them to make any formula encountered meaningful. The Grassmann algebra $\Lambda$ can be decomposed into its even and odd parts
\begin{equation}
\Lambda=\Lambda_{even}+\Lambda_{odd}.
\end{equation}
In the context of supersymmetry, the spaces $\Lambda$ and/or $\Lambda_{even}$ replace the field of complex numbers. The elements of $\Lambda$ are called supernumbers while elements of its even or odd parts are called even or odd supernumbers respectively. The Grassmann algebra can also be decomposed as
\begin{equation}
\Lambda=\Lambda_{body}+\Lambda_{soul},
\end{equation}
where
\begin{equation}
\Lambda_{body}=\Lambda^0[\xi_1,\xi_2,...]\simeq\mathbb{C},\qquad\Lambda_{soul}=\sum_{k\geqslant1}\Lambda^k[\xi_1,\xi_2,...].
\end{equation}
Here $\Lambda^0[\xi_1,\xi_2,...]$ refers to all terms that do not involve any of the generators $\xi_i$, while $\Lambda^k[\xi_1,\xi_2,...]$ refers to all terms that involve products of $k$ generators (for instance, if we have 4 generators $\xi_1,\xi_2,\xi_3,\xi_4$, then $\Lambda^2[\xi_1,\xi_2,\xi_3,\xi_4]$ refers to all terms involving $\xi_1\xi_2$, $\xi_1\xi_3$, $\xi_1\xi_4$, $\xi_2\xi_3$, $\xi_2\xi_4$ or $\xi_3\xi_4$). The bodiless elements in $\Lambda_{soul}$ are non-invertible because of the $\mathbb{Z}_0^+$-grading of the Grassmann algebra. If the number of Grassmann generators $\mathfrak{K}$ is finite, bodiless elements are nilpotent of degree at most $\mathfrak{K}$. In this paper, we assume that $\mathfrak{K}$ is arbitrarily large but finite. Our analysis is based on the global theory of supermanifolds as described in \cite{BerezinMono}-\cite{Rogers80}.

\paragraph{}Next, in our consideration, we use a $\mathbb{Z}_2$-graded complex vector space $V$ which has even basis elements $u_i$, $i=1,2,...,N$, and odd basis elements $v_\mu$, $\mu=1,2,...,N$, and construct $W=\Lambda\otimes_\mathbb{C}V$. We are interested in the even part of $W$
\begin{equation}
W_{even}=\left\lbrace\sum_ia_iu_i+\sum_\mu\underline{\alpha}_\mu v_\mu\vert a_i\in\Lambda_{even}, \underline{\alpha}_\mu\in\Lambda_{odd}\right\rbrace.
\end{equation}
Clearly, $W_{even}$ is a $\Lambda_{even}$ module which can be identified with $\Lambda^{\times N}_{even}\times\Lambda_{odd}^{\times M}$ (consisting of $N$ copies of $\Lambda_{even}$ and $M$ copies of $\Lambda_{odd}$). We associate with the original basis, consisting of $u_i$ and $v_\mu$ (although $v_\mu\in\hspace{-0.35cm}\setminus\hspace{0.2cm} W_{even}$), the corresponding functionals
\bea
E_j: W_{even}\rightarrow\Lambda_{even}: E_j\left(\sum_ia_iu_i+\sum_\mu\underline{\alpha}_\mu v_\mu\right)=a_j,\\
\Upsilon_\nu: W_{even}\rightarrow\Lambda_{odd}: \Upsilon_\nu\left(\sum_ia_iu_i+\sum_\mu\underline{\alpha_\mu} v_\mu\right)=\underline{\alpha}_\nu,
\eea
and view them as the coordinates (even and odd respectively) on $W_{even}$. Any topological manifold locally diffeomorphic to a suitable $W_{even}$ is called a supermanifold.

\paragraph{}The transitions to even and odd coordinates between different charts on the supermanifold are assumed to be even- and odd-valued superanalytic or at least $G^\infty$ functions on $W_{even}$. A comprehensive definition of the classes of supersmooth functions $G^\infty$ and superanalytic functions $G^\omega$ can be found in \cite{Rogers80}, definition 2.5. We note that superanalytic functions are those that can be expanded into a convergent power series in even and odd coordinates, whereas the definition of the $G^\infty$ function is a more involved analogue of functions on manifolds. Any $G^\infty$ function can be expanded into products of odd coordinates in a Taylor-like expansion but the coefficients, being functions of even and odd coordinates, may not necessarily be analytic (see e.g. \cite{Rogers80}).

\paragraph{}The super-Minkowski space $\mathbb{R}^{(1,1\vert 2)}$ can be viewed as such a supermanifold globally diffeomorphic to $\Lambda_{even}^{\times 2}\times \Lambda_{odd}^{\times 2}$ with even light-cone coordinates $x_+,x_-$ and odd coordinates $\theta^+,\theta^-$. Here $x_+$ and $x_-$ are linear combinations of terms involving an even number of generators : $1,\xi_1\xi_2,\xi_1\xi_3,\xi_1\xi_4,...,\xi_2\xi_3,\xi_2\xi_4,...,\xi_1\xi_2\xi_3\xi_4,...$ On the other hand, $\theta^+$ and $\theta^-$ are linear combinations of terms involving an odd number of generators : $\xi_1,\xi_2,\xi_3,\xi_4,...,\xi_1\xi_2\xi_3,\xi_1\xi_2\xi_4,\xi_1\xi_3\xi_4,\xi_2\xi_3\xi_4,...$ The supersymmetry transformation (\ref{transformation0}) and (\ref{transformation}) in the next section can be viewed as a particular change of coordinates on $\mathbb{R}^{(1,1\vert2)}$ which transforms solutions of the SUSY GW equations (\ref{3.1}) and GC equations, (\ref{GC}) respectively, into solutions of the same equation in new coordinates. A bosonic smooth superfield is a $G^\infty$ function from $\mathbb{R}^{(1,1\vert2)}$ to $\Lambda_{even}$. It can be expanded in powers of the odd coordinates $\theta^+$ and $\theta^-$ giving a decomposition in terms of even superfields
\begin{equation*}
\chi_{even}:\Lambda_{even}^{\times2}\rightarrow\Lambda_{even},
\end{equation*}
and odd superfields
\begin{equation*}
\chi_{odd}:\Lambda_{even}^{\times2}\rightarrow\Lambda_{odd}.
\end{equation*}

\paragraph{}In this paper, we use the convention that partial derivatives involving odd variables satisfy the Leibniz rule
\begin{equation}
\partial_{\theta^\pm}(hg)=(\partial_{\theta^\pm}h)g+(-1)^{\deg(h)}h(\partial_{\theta^\pm}g),
\end{equation}
where
\begin{equation}
\deg(h)=\left\lbrace\begin{array}{c}
0\mbox{ if }h\mbox{ is even,}\\
1\mbox{ if }h\mbox{ is odd,}
\end{array}\right.
\end{equation}
and the notation
\begin{equation}
f_{\theta^+\theta^-}=\partial_{\theta^-}\left(\partial_{\theta^+}f\right).
\end{equation}
The partial derivatives with respect to the odd coordinates satisfy $\partial_{\theta^i}\theta^j=\delta_i^j$ where the indices $i$ and $j$ each stand for $+$ or $-$. The operators $\partial_{\theta^\pm}$, $J_\pm$ and $D_\pm$, in equations (\ref{D}) and (\ref{J}) change the parity of a bosonic function to a fermionic function and vice versa. For example, if $\phi$ is a bosonic function, then $\partial_{\theta^+}\phi$ is an odd superfield while $\partial_{\theta^+}\partial_{\theta^-}\phi$ is an even superfield and so on. For further details see e.g. the books by Cornwell \cite{Cornwell},  DeWitt \cite{DeWitt}, Freed \cite{Freed}, Varadarajan \cite{Varadarajan} and references therein. The chain rule for a Grassmann-valued composite function $f(g(x_+))$ is
\begin{equation}
\frac{\partial f}{\partial x_+}=\frac{\partial g}{\partial x_+}\frac{\partial f}{\partial g}.
\end{equation}
The interchange of mixed derivatives (with proper respect to the ordering of odd variables) is assumed throughout. The even supernumbers, variables, fields, etc. are assumed to be elements of the even part $\Lambda_{even}$ of the underlying abstract real (complex) Grassmann ring $\Lambda$. The odd supernumbers, variables, fields, etc. lie in its odd part $\Lambda_{odd}$.

\section{Supersymmetric extension of the Gauss-Weingarten and Gauss-Codazzi equations}\label{prelim} \setcounter{equation}{0}
In this section, we derive the main elements allowing us to construct surfaces in the superspace $\mathbb{R}^{(1,1\vert2)}$. Let us consider a SUSY version of the differential equations which define surfaces in two-dimensional Minkowski space with the bosonic light-cone coordinates $x_+=\frac{1}{2}(t+x)$ and $x_-=\frac{1}{2}(t-x)$ and the fermionic (anti-commuting) variables $\theta^+$ and $\theta^-$ such that
\begin{equation}
(\theta^+)^2=(\theta^-)^2=\theta^+\theta^-+\theta^-\theta^+=0.\label{theta}
\end{equation}
Let $\mathcal{S}$ be a smooth simply connected surface in the superspace $\mathbb{R}^{(1,1\vert2)}=\lbrace(x_+,x_-,\theta^+,\theta^-)\rbrace$ which we assume is conformally parametrized in the sense that the surface $\mathcal{S}$ is given by a vector-valued bosonic superfield $F(x_+,x_-,\theta^+,\theta^-)$ satisfying conditions (\ref{gij}) specified below. Such a superfield can be decomposed in the form
\begin{equation}
\hspace{-2.5cm}F=F_m(x_+,x_-)+\theta^+\varphi_m(x_+,x_-)+\theta^-\psi_m(x_+,x_-)+\theta^+\theta^-G_m(x_+,x_-),\quad m=1,2,3\label{F}
\end{equation}
Here, the odd-valued fields $\varphi_m$ and $\psi_m$ and the even-valued fields $F_m$ and $G_m$ are the four parts of the power series with respect to $\theta^+$ and $\theta^-$ of the $m^{th}$ component of the vector superfield $F$. Let $D_+$ and $D_-$ be the covariant superspace derivatives
\be
D_\pm=\partial_{\theta^\pm}-i\theta^\pm\partial_{x_\pm}.\label{D}
\ee
Then the conformal parametrization of the surface $\mathcal{S}$ in the superspace $\mathbb{R}^{(1,1\vert2)}$ is assumed to give the following normalization of the superfield $F$
\be
\langle D_iF,D_jF\rangle=g_{ij}f,\qquad i,j=1,2\label{normalizationF}
\ee
where $f$ is a bodiless bosonic function (i.e. $f\in\Lambda_{soul}$) of $x_+$ and $x_-$ which is nilpotent of order $k$. Here the values $1$ and $2$ of the indices $i$ and $j$ stand for $+$ and $-$, respectively. The scalar product $\langle\cdot,\cdot\rangle$ in (\ref{normalizationF}) is defined in the same way as in equation (\ref{innerproduct}), taking into account the property (\ref{theta}) regarding the odd-valued variables $\theta^+$ and $\theta^-$, and taking values in the Grassmann algebra $\Lambda$. Hence the bosonic functions $g_{ij}$ of $x_+$, $x_-$, $\theta^+$ and $\theta^-$ are given by
\be
g_{11}=0,\qquad g_{12}=\frac{1}{2}e^\phi,\qquad g_{21}=-\frac{1}{2}e^\phi,\qquad g_{22}=0.\label{gij}
\ee
For the superfield $F$, given by (\ref{F}), the equations (\ref{normalizationF}) are identically satisfied for $i=j$. Indeed, in the scalar  product (\ref{innerproduct}) we have the sum of the squares of each $m^{th}$ component of the vector superfield $D_iF$. Since the square of a fermionic function vanishes, each of the terms in the scalar product is identically zero, i.e. $\langle D_iF,D_iF\rangle=0$. In the case of the mixed scalar product, the normalization imposes the condition $\langle D_+F,D_-F\rangle=\frac{1}{2}e^\phi f$. It should be noted that in equation (\ref{normalizationF}), the product $\langle D_iF,D_jF\rangle$ necessarily contains only terms involving some of the generators $\xi_1,\xi_2,\xi_3,...,\xi_i,...$ For this reason we include the bodiless function $f(x_+,x_-)$ in the normalization (\ref{normalizationF}). It is interesting to note that, by construction, the metric coefficients $g_{ij}$ satisfy the property
\be
g_{ij}=-g_{ji}.
\ee
The superfield $\phi$ is assumed to be bosonic and can be decomposed as the following power series in the odd variables $\theta^+$ and $\theta^-$
\begin{equation}
\phi =u(x_+,x_-)+\theta^+\gamma(x_+,x_-)+\theta^-\delta(x_+,x_-)+\theta^+\theta^-v(x_+,x_-).\label{Phi}
\end{equation}
Through a power expansion in $\theta^+$ and $\theta^-$ we find the exponential form
\begin{equation}
\begin{array}{l}
e^\phi=e^{u}(1+\theta^+\gamma+\theta^-\delta+\theta^+\theta^-(v-\gamma\delta)),\\
e^{-\phi}=e^{-u}(1-\theta^+\gamma-\theta^-\delta-\theta^+\theta^-(v+\gamma\delta)).
\end{array}
\end{equation}

The tangent vector superfields $D_+F$ and $D_-F$ together with the normal bosonic superfield $N(x_+,x_-,\theta^+,\theta^-)$, which can be decomposed as
\begin{equation}
\hspace{-2.5cm}N=N_m(x_+,x_-)\hspace{-0.1cm}+\hspace{-0.1cm}\theta^+\alpha_m(x_+,x_-)\hspace{-0.1cm}+\hspace{-0.1cm} \theta^-\beta_m(x_+,x_-)\hspace{-0.1cm}+\hspace{-0.1cm}\theta^+\theta^-H_m(x_+,x_-),\quad m=1,2,3\label{Nexp}
\end{equation}
form a moving frame $\Omega$ on the surface $\mathcal{S}$ in the superspace $\mathbb{R}^{(1,1\vert2)}$. Here, the even-valued fields $N_m$ and $H_m$ and the odd-valued fields $\alpha_m$ and $\beta_m$ are the four parts of the power series with respect to $\theta^+$ and $\theta^-$ of the $m^{th}$ component of the vector superfield $N$. This normal superfield $N$ has to satisfy the conditions
\begin{equation}
\langle D_iF,N\rangle=0,\qquad\langle N,N\rangle=1,\qquad i=1,2.\label{normalizationN}
\end{equation}
For any non-bodiless bosonic function $C$, one can find its inverse
\begin{equation*}
C=C_0+\theta^+C_1+\theta^-C_2+\theta^+\theta^- C_3,
\end{equation*}\begin{equation*}
C^{-1}=\frac{1}{C_0}-\theta^+\frac{C_1}{(C_0)^2}-\theta^-\frac{C_2}{(C_0)^2}-\theta^+\theta^-\left(\frac{2C_1C_2}{(C_0)^3}+\frac{C_3}{(C_0)^2}\right),
\end{equation*}
where $C_0$ is non-bodiless. So if each $N_m$ is a non-bodiless bosonic function, we can always normalize $N$ to $1$ due to the fact that
\begin{equation*}
\hspace{-2.5cm}\hat{N}=\frac{1}{\sqrt{3}}\left(N_m\hspace{-1mm}+\hspace{-1mm}\theta^+\hspace{-1mm}\alpha_m\hspace{-1mm}+\hspace{-1mm}\theta^-\hspace{-1mm} \beta_m\hspace{-1mm}+\hspace{-1mm}\theta^+\theta^-\hspace{-1mm}H_m\right)\hspace{-1mm}\left(\frac{1}{N_m}\hspace{-1mm}-\hspace{-1mm}\theta^+\hspace{-1mm}\frac{\alpha_m}{N_m^2}\hspace{-1mm}-\hspace{-1mm}\theta^-\hspace{-1mm}\frac{\beta_m}{N_m^2}\hspace{-1mm}-\hspace{-1mm}\theta^+\theta^-\hspace{-1mm}\left(\frac{2\alpha_m\beta_m}{N_m^3}\hspace{-1mm}+\hspace{-1mm}\frac{H_m}{N_m^2}\right)\right),
\end{equation*}
for $m=1,2,3,$ where $\langle \hat{N},\hat{N}\rangle=1$ holds. The covariant derivatives $D_+$ and $D_-$ have the property that they anticommute with the differential supersymmetry operators
\begin{equation}
J_+=\partial_{\theta^+}+i\theta^+\partial_{x_+},\qquad J_-=\partial_{\theta^-}+i\theta^-\partial_{x_-},\label{J}
\end{equation}
which generate the SUSY transformations
\begin{equation}
x\rightarrow x_++i\underline{\eta}_1\theta^+,\qquad \theta^+\rightarrow\theta^++i\underline{\eta}_1,\label{transformation0}
\end{equation}
and
\begin{equation}
t\rightarrow x_-+i\underline{\eta}_2\theta^-,\qquad \theta^-\rightarrow\theta^-+i\underline{\eta}_2,\label{transformation}
\end{equation}
respectively. Here $\underline{\eta}_1$ and $\underline{\eta}_2$ are odd-valued parameters. The four operators, $D_+$, $D_-$, $J_+$ and $J_-$ satisfy the anticommutation relations
\begin{equation}
\hspace{-2.5cm}\lbrace J_n,J_m\rbrace=2i\delta_{mn}\partial_{x_m},\quad \lbrace D_m,D_n\rbrace=-2i\delta_{mn}\partial_{x_m},\quad \lbrace J_m,D_n\rbrace=0,\quad m,n=1,2\label{anticommutation}
\end{equation}
where $\delta_{ij}$ is the Kronecker delta function and $\lbrace\cdot,\cdot\rbrace$ denotes the anticommutator, unless otherwise noted. Here, the values $1$ and $2$ of the indices $m$ and $n$ stand for $+$ and $-$, respectively. Therefore we have the following relations
\begin{equation}
D_\pm^2=-i\partial_\pm,\qquad J_\pm^2=i\partial_\pm.\label{relations}
\end{equation}

We now derive the SUSY version of the GW equations. Due to the normalization (\ref{normalizationF}) and (\ref{normalizationN}), the moving frame on a surface in the superspace satisfies the following GW equations
\begin{equation}
D_+\Omega=A_+\Omega,\qquad D_-\Omega=A_-\Omega,\label{3.1}
\end{equation}
where the moving frame $\Omega$ is denoted by
\begin{equation}
\Omega=\left(\ba{c}
D_+F\\
D_-F\\
N
\ea\right).
\end{equation}
Here, the first two components of $\Omega$, $D_+F$ and $D_-F$, are fermionic functions while the third component, $N$, is a bosonic function. In order to derive the SUSY GW equations, we assume that we can decompose the second-order covariant derivatives of $F$ and first-order derivatives of $N$ in terms of the tangent vectors $D_+F$ and $D_-F$ and the unit normal $N$,
\begin{equation}
\begin{array}{l}
D_jD_iF=\Gamma_{ij}^{\phantom{ij}k}D_kF+b_{ij}fN,\\
D_iN=b^k_{\phantom{k}i}D_kF+\omega_i N,
\end{array}\qquad i,j,k=1,2\label{decFN}
\end{equation}
where the coefficients $\omega_i$ and $\Gamma_{ij}^{\phantom{ij}k}$ are fermionic functions, while $b_{ij}$ and $b^k_{\phantom{k}i}$ are bosonic functions. Here, the values $1$ and $2$ of the indices $i$, $j$ and $k$ stand for $+$ and $-$, respectively. We make use of the identities
\begin{equation}
\hspace{-1.5cm}\ba{l}
0=D_i\langle N,N\rangle=\langle D_iN,N\rangle+\langle N,D_iN\rangle=2\omega_i\langle N,N\rangle=2\omega_i,\\
D_k(\frac{1}{2}e^\phi f)=D_k\langle D_+F,D_-F\rangle=\langle D_kD_+F,D_-F\rangle -\langle D_+F,D_kD_-F\rangle\\
\phantom{D_k(\frac{1}{2}e^\phi f)}=\Gamma_{1k}^{\phantom{1k}1}\langle D_+F,D_-F\rangle +\Gamma_{2k}^{\phantom{2k}2}\langle D_+F,D_-F\rangle,
\ea
\end{equation}
from which we obtain
\begin{equation}
\omega_i=0,\qquad D_kf=(\Gamma_{1k}^{\phantom{1k}1}+\Gamma_{2k}^{\phantom{2k}2}-D_k\phi)f,\label{Df}
\end{equation}
and the compatibility condition for the bodiless bosonic function $f$ in equation (\ref{Df}) is
\be
\lbrace D_+,D_-\rbrace f=\left(D_-\Gamma_{11}^{\phantom{11}1}+D_-\Gamma_{21}^{\phantom{21}2}+D_+\Gamma_{12}^{\phantom{12}1}+ D_+\Gamma_{22}^{\phantom{22}2}\right)f=0.\label{DDf}
\ee
The conformally parametrized surface $\mathcal{S}$ satisfies the normalization conditions (\ref{normalizationF}) and (\ref{normalizationN}) for the superfields $F$ and $N$, and we define the bosonic quantities $Q^+,Q^-$ and $H$ to be
\be
b_{11}=Q^+,\qquad b_{12}=\frac{1}{2}e^\phi H,\qquad b_{21}=-\frac{1}{2}e^\phi H, \qquad b_{22}=Q^-,
\ee
which gives the relations
\be
\hspace{-1.5cm}\langle D_+^2F,N\rangle=Q^+f,\qquad \langle D_-D_+F,N\rangle=\frac{1}{2}e^\phi Hf,\qquad \langle D_-^2F,N\rangle=Q^-f.\label{Q+Q-H}
\ee
Here the $b_{ij}$ have the property
\be
b_{ij}=-b_{ji},\qquad\mbox{for}\qquad i\neq j,
\ee
and are the coefficients of the second fundamental form. To obtain the bosonic functions $b^k_{\phantom{k}j}$, we make use of the relation
\be
\langle D_jD_iF,N\rangle=D_j\langle D_iF,N\rangle+\langle D_iF,D_jN\rangle=\langle D_iF,D_jN\rangle,
\ee
and by substituting $D_jN$ into its decomposition (\ref{decFN}) we get the relation
\be
\left(g_{ik}b^k_{\phantom{k}j}-b_{ij}\right)f=0.
\ee
Hence the functions $b^k_{\phantom{k}j}$ take the form
\be
\hspace{-1cm} b^1_{\phantom{1}1}=H,\qquad b^2_{\phantom{2}1}=2e^{-\phi}Q^+,\qquad b^1_{\phantom{1}2}=-2e^{-\phi}Q^-,\qquad b^2_{\phantom{2}2}=H,
\ee
up to an additional bosonic bodiless function $l\neq0$ such that $lf=0$ and where the $b^k_{\phantom{k}j}$ are the mixed coefficients of the second fundamental form. By construction, the Christoffel symbols of the second kind $\Gamma_{ij}^{\phantom{ij}k}$ have the property
\be
\Gamma_{ij}^{\phantom{ij}k}=-\Gamma_{ji}^{\phantom{ji}k}, \qquad\mbox{for }i\neq j.
\ee
Hence we define the Christoffel symbols of the first kind $\Gamma_{ijk}$ to be
\be
\Gamma_{ijk}f=\langle D_jD_iF,D_k\rangle,
\ee
so that the relations between the Christoffel symbols of the first and second kinds are
\be
(\Gamma_{ijk}-\Gamma_{ij}^{\phantom{ij}l}g_{lk})f=0,
\ee
or
\be
\Gamma_{ijk}=\Gamma_{ij}^{\phantom{ij}l}g_{lk},
\ee
up to an additional fermionic function $\zeta\neq 0$ such that $\zeta f=0$. Therefore, the Christoffel symbols of the first kind satisfy the property
\be
\Gamma_{ijk}=-\Gamma_{jik},\qquad\mbox{for } i\neq j.
\ee

\paragraph{}The matrices $A_+$ and $A_-$ of the SUSY GW equation (\ref{3.1}) are in the Bianchi form \cite{Bianchi}
\be
A_i=\left(\begin{array}{ccc}
\Gamma_{1i}^{\phantom{1i}1} & \Gamma_{1i}^{\phantom{1i}2} & b_{1i}f \\
\Gamma_{2i}^{\phantom{2i}1} & \Gamma_{2i}^{\phantom{2i}2} & b_{2i}f \\
b^1_{\phantom{1}i} & b^2_{\phantom{2}i} & 0
\end{array}\right),\qquad i=1,2,
\ee
and therefore the GW equations (\ref{3.1}) take the form
\be
\hspace{-2.5cm}\ba{cc}
D_+\Omega=A_+\Omega, & D_-\Omega=A_-\Omega,\\
A_+=\left(\ba{ccc}
R^+ & R^- & Q^+f \\
-S^+ & -S^- & -\frac{1}{2}e^\phi Hf \\
H & 2e^{-\phi}Q^+ & 0
\ea\right), & A_-=\left(\ba{ccc}
S^+ & S^- & \frac{1}{2}e^\phi Hf \\
T^+ & T^- & Q^-f \\
-2e^{-\phi}Q^- & H  & 0
\ea\right),\label{GW}
\ea
\ee
where we define the fermionic functions $R^+=\Gamma_{11}^{\phantom{11}1}$, $R^-=\Gamma_{11}^{\phantom{11}2}$, $S^+=\Gamma_{12}^{\phantom{12}1}$, $S^-=\Gamma_{12}^{\phantom{12}2}$, $T^+=\Gamma_{22}^{\phantom{22}1}$ and $T^-=\Gamma_{22}^{\phantom{22}2}$. The compatibility condition of the GW equations (\ref{GW}) is
\begin{equation*}
\hspace{-2.5cm}\ba{l}
\lbrace D_+,D_-\rbrace\Omega=D_+(A_-\Omega)+D_-(A_+\Omega),\\
\phantom{\lbrace D_+,D_-\rbrace\Omega}=D_+A_-\Omega+\left(\ba{ccc}
-R^+ & -R^- & Q^+f \\
S^+ & S^- & -\frac{1}{2}e^\phi Hf \\
H & 2e^{-\phi}Q^+ & 0
\ea\right)D_+\Omega\\
\\
\phantom{\lbrace D_+,D_-\rbrace\Omega}+D_-A_+\Omega+\left(\ba{ccc}
-S^+ & -S^- & \frac{1}{2}e^\phi Hf \\
-T^+ & -T^- & Q^-f \\
-2e^{-\phi}Q^- & H  & 0
\ea\right)D_-\Omega\\
\phantom{\lbrace D_+,D_-\rbrace\Omega}=D_+A_-\Omega-EA_-ED_+\Omega+D_-A_+\Omega-EA_+ED_-\Omega.
\ea
\end{equation*}
So we have
\be
D_+A_-+D_-A_+-\lbrace EA_+,EA_-\rbrace=0,\label{ZCC}
\ee
where
\begin{equation}
E=\pm\left(\begin{array}{ccc}
1&0&0\\
0&1&0\\
0&0&-1
\end{array}\right).
\end{equation}
The matrices $A_+$ and $A_-$ of the GW equations can also be written in the Bianchi form using matrix subblock notation
\begin{equation}
A_+=\left(\begin{array}{cc|c}
R^+ & R^- & Q^+f \\
-S^+ & -S^- & -\frac{1}{2}e^\phi Hf \\
\hline H & 2e^{-\phi}Q^+ & 0
\end{array}\right)=\left(\begin{array}{c|c}
A_f^+&I_{b_1}^+\\
\hline I_{b_2}^+&0
\end{array}\right),\label{A+}
\end{equation}
\begin{equation}
A_-=\left(\begin{array}{cc|c}
S^+ & S^- & \frac{1}{2}e^\phi Hf \\
T^+ & T^- & Q^-f \\
\hline-2e^{-\phi}Q^- & H  & 0
\end{array}\right)=\left(\begin{array}{c|c}
A_f^-&I_{b_1}^-\\
\hline I_{b_2}^-&0
\end{array}\right),\label{A-}
\end{equation}
where $A_f^+$ and $A_f^-$ are $2\times2$ matrices with fermionic entries, $I_{b_1}^+$ and $I_{b_1}^-$ are two-component column vectors with bosonic entries, and $I_{b_2}^+$ and $I_{b_2}^-$ are two-component row vectors with bosonic entries.

Indeed, let us consider a moving frame $\Psi=(\psi_f,\psi_b)$ where $\psi_f$ is a two-component fermionic vector and $\psi_b$ is a bosonic scalar. From the GW equation for the moving frame $\Omega$, with the matrices given by (\ref{A+}) and (\ref{A-}), we obtain
\begin{equation}
D_+\Psi=A_+\Psi,\qquad D_-\Psi=A_-\Psi.
\end{equation}
The compatibility conditions for $\psi_f$ and $\psi_b$ lead us to the four equations
\begin{equation}
\begin{array}{l}
D_+A^-_f+D_-A^+_f+I_{b_1}^-I_{b_2}^++I_{b_1}^+I_{b_2}^--\lbrace A_f^+,A_f^-\rbrace=0,\\
-A_f^-I_{b_1}^++D_+I_{b_1}^-+I_{b_1}^-\eta_f^+-A_f^+I_{b_1}^-+D_-I_{b_1}^++I_{b_1}^+\eta_f^-=0,\\
D_+I_{b_2}^-+I_{b_2}^-A_f^+-\eta_f^-I_{b_2}^++D_-I_{b_2}^++I_{b_2}^+A_f^--\eta_f^+I_{b_2}^-=0,\\
I_{b_2}^+I_{b_1}^-+D_-\eta_f^++I_{b_2}^-I_{b_1}^++D_+\eta_f^-=0.
\end{array}\label{ZCC2}
\end{equation}
The ZCC corresponding to the equations (\ref{ZCC2}) is an equivalent matrix form of (\ref{ZCC}).

The zero curvature condition (\ref{ZCC}) leads us to the SUSY GC equations which consist of the following six linearly independent equations for the matrix components
\be
\hspace{-2.5cm}\ba{ll}
(i) & D_-R^++D_+T^-+D_+S^+-D_-S^-=0, \\
(ii) & D_-R^+-R^-T^++D_+S^++S^-S^++\frac{1}{2}H^2e^\phi f-2Q^+Q^-e^{-\phi}f=0, \\
(iii) & Q^+T^--R^-Q^-+D_-Q^+-Q^+D_-\phi+\frac{1}{2}e^\phi D_+H=0, \\
(iv) & Q^-R^+-T^+Q^++D_+Q^--Q^-D_+\phi-\frac{1}{2}e^\phi D_-H=0, \\
(v) & D_-R^--S^+R^--R^-T^--R^+S^-+D_+S^-+2Q^+Hf=0, \\
(vi) & D_+T^++S^-T^+-T^+R^++T^-S^+-D_-S^++2Q^-Hf=0.
\ea\label{GC}
\ee
The Grassmann-valued partial differential equations (\ref{GC}) involve eleven dependent functions of the independent variables $x_+$, $x_-$, $\theta^+$ and $\theta^-$  including the four bosonic functions $\phi$, $H$, $Q^\pm$ and the six fermionic functions $R^\pm$, $S^\pm$ , $T^\pm$ together with one dependent bodiless bosonic function $f$ of $x_+$ and $x_-$. It is interesting to note that the equation (\ref{GC}.i) is the compatibility condition of the function $f$ given in equation (\ref{DDf}). 
Under the above assumptions we obtain the following result.

\begin{proposition}[Structural SUSY equations for a moving frame on a surface]
For any bosonic superfields $F(x_+,x_-,\theta^+,\theta^-)$ and $N(x_+,x_-,\theta^+,\theta^-)$ satisfying the normalization conditions (\ref{normalizationF}), (\ref{gij}), (\ref{normalizationN}) and (\ref{Q+Q-H}), the moving frame $\Omega=(D_+F,D_-F,N)^T$ on a surface immersed in the superspace $\mathbb{R}^{(1,1\vert2)}$ satisfies the SUSY GW equations (\ref{GW}). The ZCC (\ref{ZCC}), which is the compatibility condition of the SUSY GW equations (\ref{GW}) expressed in terms of the matrices $A_+$ and $A_-$, is equivalent to the SUSY GC equations (\ref{GC}).
\end{proposition}

\section{Geometric aspects of conformally parametrized supersymmetric surfaces}\setcounter{equation}{0}
In this section, we discuss certain aspects of Grassmann variables in conjunction with differential geometry and supersymmetry analysis. We begin by defining the following differential superspace fermionic operators
\begin{equation}
d_\pm=\frac{1}{2}\left[d\theta^\pm+i dx_\pm\partial_{\theta^\pm}\right],
\end{equation}
where $d_+$ and $d_-$ are the infinitesimal displacements in the direction of $D_+$ and $D_-$, respectively. These operators are anticommuting, i.e. $\lbrace d_+,d_-\rbrace=0$. For SUSY conformally parametrized surfaces, the first fundamental form is given by
\bea
\hspace{-1.5cm}I=\left\langle\left(\begin{array}{cc}
d_+ & d_-
\end{array}\right),\left(\begin{array}{cc}
d_+ & d_-
\end{array}\right)\left(\begin{array}{cc}
\langle D_+F,D_+F\rangle &\langle D_-F,D_+F\rangle\\
\langle D_+F,D_-F\rangle &\langle D_-F,D_-F\rangle
\end{array}\right)\right\rangle.\label{Iform}
\eea
Making use of the normalization of the tangent vectors $D_\pm F$ given by (\ref{normalizationF}), we have
\begin{eqnarray}
\hspace{-2,5cm}I=\left\langle\left(\begin{array}{cc}
d_+ & d_-
\end{array}\right),\left(\begin{array}{cc}
d_+ & d_-
\end{array}\right)\left(\begin{array}{cc}
0&-\frac{1}{2}fe^\phi\\
\frac{1}{2}fe^\phi&0
\end{array}\right)\right\rangle=-\frac{1}{2} fd_-d_+e^\phi+\frac{1}{2} fd_+d_-e^\phi\nonumber\\
\hspace{-2,5cm}\phantom{I}= fd_+d_-e^\phi=2fd_+d_-g_{12}.\label{R}
\end{eqnarray}
In this SUSY case, the discriminant is defined to be
\begin{equation}
g=g_{11}g_{22}-g_{12}g_{21}=\frac{1}{4}e^{2\phi}.
\end{equation}
The elements of the inverse metric form defined as
\be
g_{ij}g^{jk}=\delta_i^k,\qquad \left(\ba{cc}
g_{11} & g_{21} \\
g_{12} & g_{22}
\ea\right)\left(\ba{cc}
g^{11} & g^{21} \\
g^{12} & g^{22}
\ea\right)=\left(\ba{cc}
1 & 0 \\
0 & 1
\ea\right),
\ee
are
\begin{equation}
g^{11}=g^{22}=0,\qquad g^{12}=-g^{21}=-2e^{-\phi} .
\end{equation}
The SUSY version of the second fundamental form is
\bea
\hspace{-1.5cm}I\hspace{-0.1cm}I=\left\langle\left(\begin{array}{cc}
d_+ & d_-
\end{array}\right),\left(\begin{array}{cc}
d_+ & d_-
\end{array}\right)\left(\begin{array}{cc}
\langle D_+^2F,N\rangle & \langle D_+D_-F,N\rangle\\
\langle D_-D_+F,N\rangle &\langle D_-^2F,N\rangle
\end{array}\right)\right\rangle.\label{IIform}
\eea
By virtue of the normalization of the tangent vectors $D_\pm F$ and the unit normal $N$, given by (\ref{normalizationF}) and (\ref{normalizationN}) respectively, we have
\bea
\hspace{-1.5cm}I\hspace{-0.1cm}I&=\left\langle\left(\begin{array}{cc}
d_+ & d_-
\end{array}\right),\left(\begin{array}{cc}
d_+ & d_-
\end{array}\right)\left(\begin{array}{cc}
Q^+f&-\frac{1}{2}He^\phi f\\
\frac{1}{2}He^\phi f &Q^-f
\end{array}\right)\right\rangle\nonumber\\
\hspace{-1.5cm}&=f(d_+^2Q^+-\frac{1}{2}d_-d_+(He^\phi) +\frac{1}{2} d_+d_-(He^\phi)+d_-^2Q^-)\nonumber\\
\hspace{-1.5cm}&=f(d_+^2Q^++d_+d_-(He^\phi) +d_-^2Q^-)=f\left(d_+^2b_{11}+2d_+d_-b_{12}+d_-^2b_{22}\right).\label{S}
\eea
In order to compute the first and second fundamental forms, we have assumed that $(d\theta^j\_\hspace{-0.15cm}\shortmid\partial_{\theta^i})=0$, for $i, j=1,2$. In the SUSY case, the discriminant is defined to be
\begin{equation}
b=b_{11}b_{22}-b_{12}b_{21}=Q^+Q^-+\frac{1}{4}H^2e^{2\phi}.
\end{equation}
From the first and second fundamental forms, we can determine the SUSY version of the Gaussian curvature $\mathcal{K}$ and the mean curvature $H$ as follows
\begin{equation}
\hspace{-2,5cm}\mathcal{K}=\det(SR^{-1})=\frac{b}{g}=4Q^+Q^-e^{-2\phi}+H^2,\qquad H=\frac{1}{2}tr(SR^{-1}),
\end{equation}
where the matrices $R$ and $S$ are similar to the matrices in equations (\ref{R}) and (\ref{S}) respectively (without the function $f$) and can be written as
\begin{equation*}
\hspace{-2.5cm}R=\left(\ba{cc}
g_{11} & g_{21} \\
g_{12} & g_{22}
\ea\right)=\frac{1}{2}e^{\phi}\left(\begin{array}{cc}
0&-1\\
1&0
\end{array}\right),\quad S=\left(\ba{cc}
b_{11} & b_{21} \\
b_{12} & b_{22}
\ea\right)=\left(\begin{array}{cc}
Q^+&-\frac{1}{2}He^\phi\\
\frac{1}{2}e^\phi H&Q^-
\end{array}\right).
\end{equation*}
The determinant is well-defined for $R$ and $S$ since both $R$ and $S$ are bosonic-valued matrices.
Based on the SUSY version of the GC equations (\ref{GC}) we can provide a SUSY analogue of the Bonnet Theorem.
\begin{proposition}[Supersymmetric extension of the Bonnet theorem]
Given a \\SUSY conformal metric
\begin{equation}
M=fd_+d_-e^\phi 
\end{equation}
of a conformally parametrized surface $\mathcal{S}$, the Hopf differentials $d_\pm^2Q^\pm $ and a mean curvature function $H$ defined on a Riemann surface $\mathcal{R}$ satisfying the GC equations (\ref{GC}), there exists a vector-valued bosonic immersion function
\begin{equation}
F=(F_1,F_2,F_3):\tilde{\mathcal{R}}\rightarrow\mathbb{R}^{(1,1\vert2)},
\end{equation}
with the fundamental forms
\begin{equation}
I= fd_+d_-e^\phi,\qquad I\hspace{-0.1cm}I=f(d_+^2Q^++ d_+d_-(He^\phi)+d_-^2Q^-),
\end{equation}
where $\tilde{\mathcal{R}}$ is the universal covering of the Riemann surface $\mathcal{R}$ and $\mathbb{R}^{(1,1\vert2)}$ is the superspace. The immersion function $F$ is unique up to affine transformations in the superspace $\mathbb{R}^{(1,1\vert2)}$.
\end{proposition}

\paragraph{}The proof of this proposition is analogous to that given in \cite{Bonnet}. Note that it is straightforward to construct surfaces on the superspace $\mathbb{R}^{(1,1\vert2)}$ related to integrable equations. However, it is nontrivial to identify those surfaces which have an invariant geometrical characterization. A list of such surfaces is known in the classical case \cite{Bob} but, to our knowledge, an identification of such surfaces is an open problem in the case of surfaces immersed in the superspace.

\section{Symmetries of the supersymmetric Gauss-Codazzi equations}\setcounter{equation}{0}
A symmetry supergroup $G$ of a SUSY system is a (local) supergroup of transformations acting on a Cartesian product of supermanifolds $X\times \mathcal{U}$, where $X$ is the space of four independent variables $(x_+,x_-,\theta^+,\theta^-)$ and $\mathcal{U}$ is the space of eleven dependent superfields $(\phi,H,Q^+,Q^-,R^+,R^-,^+,S^-,T^+,T^-,f)$. The action of the group $G$ on the functions $\phi$, $H$, $Q^+$, $Q^-$, $R^+$, $R^-$, $S^+$, $S^-$, $T^+$, $T^-$ and $f$ of $(x_+,x_-,\theta^+,\theta^-)$ maps solutions of (\ref{GC}) into solution of (\ref{GC}). Strictly speaking, the bodiless bosonic function $f$ introduced in (\ref{normalizationF}) depends only on $x_+$ and $x_-$, which has to be taken into consideration when we perform the symmetry reductions. If we assume that $G$ is a Lie supergroup as described in \cite{Winternitz},\cite{Clarkson},\cite{Kac}, one can associate it with its Lie superalgebra $\mathfrak{g}$ whose elements are infinitesimal symmetries of (\ref{GC}). The SUSY GC equations (\ref{GC}) are invariant under the Lie superalgebra $\mathfrak{g}$ generated by the following eight infinitesimal vector fields
\be
\hspace{-2.5cm}\ba{l}
C_0= H\partial_H+Q^+\partial_{Q^+}+Q^-\partial_{Q^-}-2f\partial_f,\\
K_0= -H\partial_H+Q^+\partial_{Q^+}+Q^-\partial_{Q^-}+2\partial_\phi,\\
K_1=-2x_+\partial_{x_+}-\theta^+\partial_{\theta^+}+R^+\partial_{R^+}+2R^-\partial_{R^-}+S^-\partial_{S^-}-T^+\partial_{T^+}+2Q^+\partial_{Q^+}+\partial_\phi,\\
K_2=-2x_-\partial_{x_-}-\theta^-\partial_{\theta^-}-R^-\partial_{R^-}+S^+\partial_{S^+}+2T^+\partial_{T^+}+ T^-\partial_{T^-}+2Q^-\partial_{Q^-}+\partial_\phi,\\
P_+=\partial_{x_+},\qquad\qquad\qquad\qquad P_-=\partial_{x_-},\\
J_+=\partial_{\theta^+}+i\theta^+\partial_{x_+},\qquad\qquad J_-=\partial_{\theta^-}+i\theta^-\partial_{x_-}.
\ea\label{gen}
\ee
The generators $P_+$ and $P_-$ represent translations in the bosonic variables $x_+$ and $x_-$ while $K_1$, $K_2$, $K_0$ and $C_0$ generate dilations on both even and odd variables. In addition, we recover the supersymmetry transformations $J_+$ and $J_-$ which were identified previously in (\ref{J}). In order to determine this superalgebra of infinitesimal symmetries, we have made use of the theory described in the book by P. Olver \cite{Olver}.

The commutation (anticommutation in the case of two fermionic operators) relations of the superalgebra $\mathfrak{g}$ of the SUSY GC equations (\ref{GC}) are given in Table~1 for the case $D_\pm f\neq0$. 

\begin{table}[h!]
\centering
\caption{Commutation table for the Lie superalgebra $\mathfrak{g}$ spanned by\\ the vector fields (\ref{gen}). In the case of two fermionic generator $J_+$\\ and/or $J_-$ we have anticommutation rather than commutation.}
\begin{tabular}{c|c|c|c|c|c|c|c|c|}
 &$K_1$&$P_+$&$J_+$&$K_2$&$P_-$&$J_-$&$K_0$&$C_0$\\
\hline$K_1$&$0$ &$2P_+$ &$J_+$ &$0$ &$0$ &$0$ &$0$ &$0$ \\
\hline$P_+$&$-2P_+$ &$0$ &$0$ &$0$ &$0$ &$0$ &$0$ &$0$ \\
\hline$J_+$&$-J_+$ &$0$ &$2iP_+$ &$0$ &$0$ &$0$ &$0$ &$0$ \\
\hline$K_2$&$0$ &$0$ &$0$ &$0$ &$2P_-$ &$J_-$ &$0$ &$0$ \\
\hline$P_-$&$0$ &$0$ &$0$ &$-2P_-$ &$0$ &$0$ &$0$ &$0$ \\
\hline$J_-$&$0$ &$0$ &$0$ &$-J_-$ &$0$  &$2iP_-$ &$0$ &$0$ \\
\hline$K_0$&$0$ &$0$ &$0$ &$0$ &$0$ &$0$ &$0$ &$0$ \\
\hline$C_0$&$0$ &$0$ &$0$ &$0$ &$0$ &$0$ &$0$ &$0$ \\
\hline
\end{tabular}
\end{table}
\noindent The Lie superalgebra $\mathfrak{g}$ can be decomposed into the following combination of direct and semi-direct sums
\begin{equation}
\mathfrak{g}=\lbrace\lbrace K_1\rbrace\sdir\lbrace P_+,J_+\rbrace\rbrace \oplus\lbrace\lbrace K_2\rbrace\sdir\lbrace P_-,J_-\rbrace\rbrace\oplus\lbrace K_0\rbrace\oplus\lbrace C_0\rbrace.\label{cla}
\end{equation}
In equation (\ref{cla}) the braces $\lbrace\cdot,...,\cdot\rbrace$ denote the set of generators listed in (\ref{gen}). It should be noted that $K_0$ and $C_0$ constitute the center of the Lie superalgebra $\mathfrak{g}$.

\section{One-dimensional subalgebras of the symmetry superalgebra of the supersymmetric Gauss-Codazzi equations}\setcounter{equation}{0}
In this section, we perform a classification of the one-dimensional subalgebras of the Lie superalgebra of infinitesimal transformations $\mathfrak{g}$ into conjugacy classes under the action of the Lie supergroup $\exp(\mathfrak{g})$ generated by (\ref{gen}). The significance of such a classification resides in the fact that conjugate subgroups necessarily lead to invariant solutions which are equivalent in the sense that they can be transformed from one to the other by a suitable symmetry. Therefore, it is not necessary to compute reductions with respect to algebras which are conjugate to each other.

The significance of the algebra $\mathfrak{g}$ resides in the following facts. It would be inconsistent to consider the $\mathbb{R}$ or $\mathbb{C}$ span of the generators (\ref{gen}) because we multiply the odd generators $J_+$ and $J_-$ by the odd parameters $\underline{\mu}$ and $\underline{\eta}$ respectively in equation (\ref{Li}). Therefore, one is naturally led to consider $\mathfrak{g}$ which is a supermanifold in the sense presented in section 2. This means that $\mathfrak{g}$ contains sums of any even combinations of $P_+$, $P_-$, $K_1$, $K_2$, $K_0$ and $C_0$ (i.e. multiplied by even parameters in $\mathfrak{g}$ including real or complex numbers) and odd combinations of $J_+$ and $J_-$ (i.e. multiplied by odd parameters in $\Lambda_{odd}$). At the same time $\mathfrak{g}$ is a $\Lambda_{even}$ Lie module. This fact can lead to the following complication. For a given $X\in\mathfrak{g}$, the subalgebras $\mathfrak{X}$ and $\mathfrak{X}'$ spanned by $X$ and $X'=aX$ with $a\in\Lambda_{even}\backslash\mathbb{C}$ are not isomorphic in general, i.e. $\mathfrak{X}'\subset\mathfrak{X}$.

Note that the subalgebras obtained from other ones through multiplication by bodiless elements of $\Lambda_{even}$ do not provide us with anything new for the purpose of symmetry reduction. These subalgebras may allow for more freedom in the choice of invariants, but we then encounter the problem of non-standard invariants \cite{GH11},\cite{GHS09},\cite{GHS11} which are discussed at the end of this section. Note also that it does not appear to be particularly useful to consider a subalgebra of the form e.g. $\lbrace P_++\underline{\eta}_1\underline{\eta}_2 P_-\rbrace$. 

In what follows, we will assume throughout the computation of the non-isomorphic one-dimensional subalgebras that the nonzero bosonic parameters are invertible (i.e. behave essentially like ordinary real or complex numbers.) In order to classify the Lie superalgebra (\ref{cla}) under the action of the supergroup generated by $\mathfrak{g}$, we make use of the techniques for classifying direct and semi-direct sums of algebras described in \cite{Winternitz} and generalize them to superalgebras involving both even and odd generators. In the case of direct sums, we use the Goursat twist method generalized to the case of a superalgebra. Here the superalgebra (\ref{cla}) contains two isomorphic copies of the 3-dimensional algebra $\mathfrak{g}_1=\lbrace\lbrace K_1\rbrace \sdir\lbrace P_+,J_+\rbrace\rbrace$ (the other copy being $\mathfrak{g}_2=\lbrace\lbrace K_2\rbrace \sdir\lbrace P_-,J_-\rbrace\rbrace$) together~with the one-dimensional algebras $\lbrace K_0\rbrace$ and $\lbrace C_0\rbrace$ which constitute the center of the Lie superalgebra $\mathfrak{g}$. This fact allows us to adapt the classification for 3-dimensional algebras as described in \cite{Patera}. So we begin our classification by considering the twisted one-dimensional subalgebras of $\mathfrak{g}_1\oplus\mathfrak{g}_2$. Under the action of a one-parameter group generated by the vector field
\begin{equation}
X=\alpha K_1+\beta P_++\underline{\eta}J_++\delta K_2+\lambda P_-+\underline{\rho}J_-,\label{7.3}
\end{equation}
where $\alpha,\beta,\delta,\lambda\in\Lambda_{even}$ and $\underline{\eta},\underline{\rho}\in\Lambda_{odd}$, the one-dimensional subalgebra
\begin{equation*}
Y=P_++aP_-,\qquad a\in\Lambda_{even}
\end{equation*}
transforms under the Baker-Campbell-Hausdorff formula
\begin{equation}
\hspace{-2cm}Y\rightarrow \mbox{Ad}_{\exp(X)}Y=Y+[X,Y]+\frac{1}{2!}[X,[X,Y]]+\frac{1}{3!}[X,[X,[X,Y]]]+...\label{BCH}
\end{equation}
to $e^{-2\alpha}P_++e^{-2\delta}aP_-$. Hence we get that $\lbrace P_++aP_-\rbrace$ is isomorphic to $\lbrace P_++e^{2\alpha-2\delta}aP_-\rbrace$. By a suitable choice of $\alpha$ and $\delta$, the factor $e^{2\alpha-2\delta}a$ can be re-scaled to either $1$ or $-1$. Hence, we obtain a twisted subalgebra $L_{14}=\lbrace P_++\epsilon P_-,\epsilon=\pm1\rbrace$.

As another example, consider a twisted subalgebra of the form $\lbrace P_++aK_2,~a\neq0\rbrace$, where $a\in\Lambda_{even}$. Through the Baker-Campbell-Hausdorff formula (\ref{BCH}), the vector field $Y=K_2+aP_+$ transforms (through the vector field $X$ given in (\ref{7.3})) to
\begin{equation}
e^XYe^{-X}=K_2+e^{-2\alpha}aP_+-\frac{\lambda}{\delta}(e^{-2\delta}-1)P_--\frac{1}{\delta}(e^{-\delta}-1)\underline{\rho}J_-.\label{BCH2}
\end{equation}
Through a suitable choice of $\underline{\lambda}$ and $\underline{\rho}$, the last two terms of (\ref{BCH2}) can be eliminated, so we obtain the twisted subalgebra $L_{13}=\lbrace K_2+\epsilon P_+,\epsilon=\pm1\rbrace$. Continuing the classification in an analogous way, we obtain the following list of one-dimensional subalgebras
\be
\hspace{-1.2cm}\ba{rl}
L_1=\lbrace K_1\rbrace, & L_{26}=\lbrace K_1+aK_0\rbrace, \\
L_2=\lbrace P_+\rbrace, & L_{27}=\lbrace K_0+\epsilon P_+\rbrace, \\
L_3=\lbrace \underline{\mu}J_+\rbrace, & L_{28}=\lbrace K_0+\underline{\mu}J_+\rbrace, \\
L_4=\lbrace P_++\underline{\mu}J_+\rbrace, & L_{29}=\lbrace K_0+\epsilon P_++\underline{\mu}J_+\rbrace, \\
L_5=\lbrace K_2\rbrace, & L_{30}=\lbrace K_2+aK_0\rbrace, \\
L_6=\lbrace P_-\rbrace, & L_{31}=\lbrace K_0+\epsilon P_-\rbrace, \\
L_7=\lbrace \underline{\nu}J_-\rbrace, & L_{32}=\lbrace K_0+\underline{\nu}J_-\rbrace, \\
L_8=\lbrace P_-+\underline{\nu}J_-\rbrace, & L_{33}=\lbrace K_0+\epsilon P_-+\underline{\nu}J_-\rbrace, \\
L_9=\lbrace K_1+aK_2\rbrace, & L_{34}=\lbrace K_1+aK_2+bK_0\rbrace, \\
L_{10}=\lbrace K_1+\epsilon P_-\rbrace, & L_{35}=\lbrace K_1+aK_0+\epsilon P_-\rbrace, \\
L_{11}=\lbrace K_1+\underline{\nu}J_-\rbrace, & L_{36}=\lbrace K_1+aK_0+\underline{\nu}J_-\rbrace, \\
L_{12}=\lbrace K_1+\epsilon P_-+\underline{\nu}J_-\rbrace, & L_{37}=\lbrace K_1+aK_0+\epsilon P_-+\underline{\nu}J_-\rbrace,\\
L_{13}=\lbrace K_2+\epsilon P_+\rbrace, & L_{38}=\lbrace K_2+aK_0+\epsilon P_+\rbrace, \\
L_{14}=\lbrace P_++\epsilon P_-\rbrace, & L_{39}=\lbrace K_0+\epsilon_1 P_++\epsilon_2 P_-\rbrace, \\
L_{15}=\lbrace P_++\underline{\nu}J_-\rbrace, & L_{40}=\lbrace K_0+\epsilon P_++\underline{\nu}J_-\rbrace, 
\ea\ee
\setcounter{equation}{3}
\be\hspace{-2.5cm}\ba{rl}
L_{16}=\lbrace P_++\epsilon P_-+\underline{\nu}J_-\rbrace, & L_{41}=\lbrace K_0+\epsilon_1 P_++\epsilon_2 P_-+\underline{\nu}J_-\rbrace, \\
L_{17}=\lbrace K_2+\underline{\mu}J_+\rbrace, & L_{42}=\lbrace K_2+aK_0+\underline{\mu}J_+\rbrace, \\
L_{18}=\lbrace P_-+\underline{\mu}J_+\rbrace, & L_{43}=\lbrace K_0+\epsilon P_-+\underline{\mu}J_+\rbrace, \\
L_{19}=\lbrace \underline{\mu}J_++\underline{\nu}J_-\rbrace, & L_{44}=\lbrace K_0+\underline{\mu}J_++\underline{\nu}J_-\rbrace, \\
L_{20}=\lbrace P_-+\underline{\mu}J_++\underline{\nu}J_-\rbrace, & L_{45}=\lbrace K_0+\epsilon P_-+\underline{\mu}J_++\underline{\nu}J_-\rbrace, \\
L_{21}=\lbrace K_2+\epsilon P_++\underline{\mu}J_+\rbrace, & L_{46}=\lbrace K_2+aK_0+\epsilon P_++\underline{\mu}J_+\rbrace, \\
L_{22}=\lbrace P_++\epsilon P_-+\underline{\mu}J_+\rbrace, & L_{47}=\lbrace K_0+\epsilon_1 P_++\epsilon_2P_-+\underline{\mu}J_+\rbrace, \\
L_{23}=\lbrace P_++\underline{\mu}J_++\underline{\nu}J_-\rbrace, & L_{48}=\lbrace K_0+\epsilon P_++\underline{\mu}J_++\underline{\nu}J_-\rbrace, \\
L_{24}=\lbrace P_++\epsilon P_-+\underline{\mu}J_++\underline{\nu}J_-\rbrace, & L_{49}=\lbrace K_0+\epsilon_1P_++\epsilon_2P_-+\underline{\mu}J_++\underline{\nu}J_-\rbrace, \\
L_{25}=\lbrace K_0\rbrace, & 
\ea\label{Li}
\ee
where $\epsilon$ is either $\pm1$, the parameters $a$ and $b$ are non-zero bosonic constants and $\underline{\mu}$ and $\underline{\nu}$ are fermionic constants. In addition to these $49$ subalgebras, we also have the subalgebras obtained when, for each of the subalgebras from $L_{25}$ to $L_{49}$, the vector field $K_0$ is replaced by the vector field $C_0$ and/or by the vector $K_0+mC_0$ for an arbitrary non-zero bosonic constant $m$. For each subalgebra $L_k$ (where $25\leq k\leq49$) involving $K_0$, the equivalent subalgebra where $K_0$ is replaced by $C_0$ is labelled $L_k'$, and the equivalent subalgebra where $K_0$ is replaced by $K_0+mC_0$ is labelled $L_k''$. Since this would lengthen our list considerably we do not list them here. These representative subalgebras allow us to determine invariant solutions of the SUSY GC equations (\ref{GC}) using the symmetry reduction method.

In addition, we note that for certain one-dimensional subalgebras (e.g. $L_{25}$, $L_{25}'$ and $L_{25}''$), the invariants have a non-standard form in the sense that they do not lead to standard reductions or invariant solutions. Such non-standard invariants were found by the authors for several other SUSY hydrodynamic-type systems, including the SUSY polytropic gas dynamics \cite{GH11}, the SUSY sine-Gordon equation \cite{GHS09} and SUSY Klein-Gordon polynomial equations \cite{GHS11}.

\section{Invariant solutions of the supersymmetric GC equations}\setcounter{equation}{0}
We now make use of the symmetry reduction method (SRM) in order to obtain invariant solutions of the GC equations (\ref{GC}). For each subalgebra, the superfields $\mathcal{U}=(\phi,H,Q^+,Q^-,R^+,R^-,^+,S^-,T^+,T^-,f)$ are expanded in terms of the various invariants. The dependence of the components of $\mathcal{U}$ on each odd variable $\theta^+$ or $\theta^-$ must be at most linear as the odd variables satisfy (\ref{theta}). Substituting this decomposition into the GC equations (\ref{GC}), we obtain reduced partial differential equations for the superfields $\mathcal{U}$ which in turn lead to systems of differential constraints between their component even and odd functions. For instance, if the invariants are given by the bosonic symmetry variable $\xi$ and the fermionic symmetry variables $\eta$ and $\sigma$ (which are expressed in terms of $\theta^+$ and $\theta^-$, respectively), then $\mathcal{U}$ can be decomposed into the form
\begin{equation}
\begin{array}{l}
Q^\pm=q_0^\pm(\xi)+\eta q^\pm_1(\xi)+\sigma q_2^\pm(\xi)+\eta\sigma q_3^\pm(\xi),\\
R^\pm=r_0^\pm(\xi)+\eta r^\pm_1(\xi)+\sigma r_2^\pm(\xi)+\eta\sigma r_3^\pm(\xi),\\
S^\pm=s_0^\pm(\xi)+\eta s^\pm_1(\xi)+\sigma s_2^\pm(\xi)+\eta\sigma s_3^\pm(\xi),\\
T^\pm=t_0^\pm(\xi)+\eta t^\pm_1(\xi)+\sigma t_2^\pm(\xi)+\eta\sigma t_3^\pm(\xi),\\
H=h_0(\xi)+\eta h_1(\xi)+\sigma h_2(\xi)+\eta\sigma h_3(\xi),\\
\phi=\phi_0(\xi)+\eta \phi_1(\xi)+\sigma \phi_2(\xi)+\eta\sigma \phi_3(\xi),\\
f=\psi(\xi),
\end{array}\label{7.2}
\end{equation}
where $q^\pm_0,q^\pm_3,r_1^\pm,r_2^\pm,s_1^\pm,s_2^\pm,t_1^\pm,t_2^\pm,h_0,h_3,\phi_0,\phi_3$ and $\psi$ are even-valued functions of $\xi$ while $q_1^\pm,q_2^\pm,r_0^\pm,r_3^\pm,s_0^\pm,s_3^\pm,t_0^\pm,t_3^\pm,h_1,h_2,\phi_1$ and $\phi_2$ are odd-valued functions of $\xi$. We now present the following three examples in order to illustrate the geometrical considerations.

\paragraph{}\textbf{1.} For the subalgebra $L_{39}=\lbrace P_++\epsilon P_-+aK_0,~\epsilon=\pm1,~a\neq0\rbrace$, the orbit of the group of the SUSY GC equations (\ref{GC}) can be parametrized as follows
\be
\ba{ll}
H=e^{-ax_+}h(\xi,\theta^+,\theta^-),&\\
Q^+=e^{ax_+}q^+(\xi,\theta^+,\theta^-),& S^+=s^+(\xi,\theta^+,\theta^-),\\
Q^-=e^{ax_+}q^-(\xi,\theta^+,\theta^-),& S^-=s^-(\xi,\theta^+,\theta^-),\\
R^+=r^+(\xi,\theta^+,\theta^-),& T^+=t^+(\xi,\theta^+,\theta^-),\\
R^-=r^-(\xi,\theta^+,\theta^-),& T^-=t^-(\xi,\theta^+,\theta^-),\\
\phi=2ax_++\varphi(\xi,\theta^+,\theta^-),& f=\psi(\xi),
\ea
\ee
where the functions $H,Q^\pm,R^\pm,S^\pm,T^\pm$ and $\phi$ are expressed in terms of the bosonic symmetry variable $\xi=x_--\epsilon x_+$ and the fermionic symmetry variables $\theta^+$ and $\theta^-$. A corresponding invariant solution is given by
\be
\hspace{-2.5cm}\ba{l}
H=e^{-ax_+}\left[h_0+\theta^+\theta^-2il_0e^\xi\right],\\
Q^+=e^{ax_+}\left[l_0e^{2\xi}+l_1e^\xi+\theta^+\theta^-\left(\frac{1}{2}ie^\xi(ah_0+\epsilon(h_0)_\xi)+l_0e^{2\xi}\varphi_1+l_1e^\xi\varphi_1\right)\right],\\
Q^-=e^{ax_+}\left[\frac{\epsilon l_0}{a\epsilon-1}+l_2e^{(1-a\epsilon)\xi}+\theta^+\theta^-\left(-\frac{1}{2}ie^\xi(h_0)_\xi+\frac{\epsilon l_0}{a\epsilon-1}\varphi_1+l_2e^{(1-a\epsilon)\xi}\varphi_1\right)\right],\\
R^-=b_1\underline{S}^+_0,\quad R^+=b_2\underline{S}^+_0,\quad S^+=\underline{S}_0^+,\quad S^-=\underline{S}_0^+,\quad T^-=b_3\underline{T}^+_0,\quad T^+=b_4\underline{S}^+_0,\\
\phi=2ax_++\xi+\theta^+\theta^-\varphi_1,\qquad f=\psi,\\
l_0=\underline{a}_0\underline{S}_0^+,\qquad\quad l_1=\underline{a}_2\underline{S}_0^+,\qquad l_2=\underline{a}_2\underline{S}_0^+,\qquad\quad h_0=\underline{c}_0\underline{S}_0^+,
\ea\label{quan2}
\ee
where $h_0$, $\varphi_1$ and $\psi$ are functions of the symmetry variable $\xi=x_--\epsilon x_+$ and where $l_0,l_1,l_2$ and $b_1,b_2,b_3,b_4$ are bosonic constants, while $\underline{S}_0^+,\underline{c}_0$ and $\underline{a}_0,\underline{a}_1,\underline{a}_2$ are fermionic constants.

The first and second fundamental forms of the surface $\mathcal{S}$ associated with (\ref{quan2}) are given by
\be
\hspace{-2.5cm}\ba{ll}
I=\psi d_+d_-\left[e^{2ax_++\xi}\left(1+\theta^+\theta^-\varphi_1\right)\right],\\
I\hspace{-0.1cm}I=\psi e^{ax_+}\left\lbrace d_+^2\left[l_0e^{2\xi}+l_1e^\xi+\theta^+\theta^-\left(\frac{1}{2}ie^\xi(ah_0+\epsilon(h_0)_\xi)+l_0e^{2\xi}\varphi_1+l_1e^\xi\varphi_1\right)\right]\right.\\
\phantom{I\hspace{-0.1cm}I=\psi e^{ax_+}}+d_+d_-\left[e^\xi\left(h_0+\theta^+\theta^-(2il_0e^\xi+h_0\varphi_1)\right)\right]\\
\left.\phantom{I\hspace{-0.1cm}I=\psi e^{ax_+}}+d_-^2\left[\frac{\epsilon l_0}{a\epsilon-1}+l_2e^{(1-a\epsilon)\xi}+\theta^+\theta^-\left(-\frac{1}{2}ie^\xi(h_0)_\xi+\frac{\epsilon l_0}{a\epsilon-1}\varphi_1+l_2e^{(1-a\epsilon)\xi}\varphi_1\right)\right]\right\rbrace.
\ea
\ee
The Gaussian curvature takes the form
\be
\hspace{-2.6cm}\ba{l}
\mathcal{K}\hspace{-0.1cm}=\hspace{-0.1cm}e^{-2ax_+}\hspace{-0.1cm}\left[h_0^2+\theta^+\theta^-4ih_0l_0e^\xi+4(l_0e^{2\xi}+l_1e^\xi)(\frac{\epsilon l_0}{a\epsilon-1}+l_2e^{(1-a\epsilon)\xi})e^{-2\xi}(1-\theta^+\theta^-2\varphi_1)\right.\\
\phantom{\mathcal{K}=e^{-2ax_+}}+4\theta^+\theta^-(l_0e^{2\xi}+l_1e^\xi)\left(-\frac{1}{2}ie^\xi(h_0)_\xi+\frac{\epsilon l_0}{a\epsilon-1}\varphi_1+l_2e^{(1-a\epsilon)\xi}\varphi_1\right)e^{-2\xi}\\
\phantom{\mathcal{K}=e^{-2ax_+}}\left.+4\theta^+\theta^-(\frac{\epsilon l_0}{a\epsilon-1}+l_2e^{(1-a\epsilon)\xi})\left(\frac{1}{2}ie^\xi(ah_0+\epsilon(h_0)_\xi)+l_0e^{2\xi}\varphi_1+l_1e^\xi\varphi_1\right)e^{-2\xi}\right].
\ea
\ee
The subalgebra of the classical GC equation (\ref{cCeq}) analogous to $L_{39}$ is $L_{1,7}'=\lbrace e_1+\epsilon e_2+ae_0,\epsilon=\pm1,a\neq0\rbrace$, whose corresponding invariant solution is given by
\be
\ba{ll}
H(z,\bar{z})=k_0v(\xi)^{-1/2}e^{a/2(\bar{z}-3z)},& Q(z,\bar{z})=\frac{1}{2}k_0v(\xi)^{1/2}e^{a/2(z+\bar{z})},\\
U(z,\bar{z})=e^{2az}v(\xi),& \bar{Q}(z,\bar{z})=\frac{1}{2}k_0v(\xi)^{1/2}e^{a/2(z+\bar{z})},
\ea
\ee
where the symmetry variable is $\xi=\bar{z}-\epsilon z$ and the function $v$ of $\xi$ satisfies the ODE
\be
v_{\xi\xi}=\frac{(v_\xi)^2}{v}+k_0^2ve^{a\xi}.
\ee
For this classical solution, the Gaussian curvature vanishes, in contrast to the SUSY case.

\paragraph{}\textbf{2.} For the subalgebra $L_{27}'=\lbrace C_0+\epsilon P_+,~\epsilon=\pm1\rbrace$ we obtain the following orbits of the group
\be
\ba{ll}
H=e^{\epsilon x_+}h(x_-,\theta^+,\theta^-),& \\
Q^+=e^{\epsilon x_+}q^+(x_-,\theta^+,\theta^-),& S^+=s^+(x_-,\theta^+,\theta^-),\\
Q^-=e^{\epsilon x_+}q^-(x_-,\theta^+,\theta^-),& S^-=s^-(x_-,\theta^+,\theta^-),\\
R^+=r^+(x_-,\theta^+,\theta^-),& T^+=t^+(x_-,\theta^+,\theta^-),\\
R^-=r^-(x_-,\theta^+,\theta^-),& T^-=t^-(x_-,\theta^+,\theta^-),\\
\phi=\varphi(x_-,\theta^+,\theta^-),& f=e^{-2\epsilon x_+}\psi(x_-),
\ea
\ee
where the symmetry variables are $x_-,\theta^+$ and $\theta^-$.~An invariant solution of the SUSY GC equations (\ref{GC}) is given by
\be
\hspace{-2.5cm}\ba{l}
H=e^{\epsilon x_+}\left[h_0(x_-)+h_1(x_-)\theta^+\theta^-\right],\\
Q^+=e^{\epsilon x_+}\left[B_0^++B_1^+\theta^+\theta^-\right]\psi(x_-),\qquad Q^-=e^{\epsilon x_+}\left[B_0^-+B_1^-\theta^+\theta^-\right]\psi(x_-),\\
R^+=0,\quad R^-=0,\quad S^+=\underline{S}_0^+,\quad S^-=a\underline{S}_0^+,\quad T^+=0,\quad T^-=0,\\
\phi=\varphi_0(x_-)+\varphi_1(x_-)\theta^+\theta^-,\qquad f=e^{-2\epsilon x_+}\psi(x_-),
\ea
\ee
where $\underline{S}_0^+$ is a fermionic constant while $a$, $B_0^\pm$ and $B_1^\pm$ are bosonic constants. The bosonic functions $h_0$ and $h_1$ obey the relations
\be
h_0=2i\epsilon(B_1^+-B_0^+\varphi_1)e^{-\varphi_0}\psi,\qquad h_1=2i\epsilon B_0^-e^{-\varphi_0}\psi.
\ee
The bosonic function $\varphi_1$ is given by
\be
\varphi_1=\frac{\epsilon}{(B_0^+)^2}(B_0^-B_1^+-B_0^+B_1^-)x_-+k_0,
\ee
where $k_0$ is a bosonic constant and the function $\varphi_0$ obeys the ODE
\be
\psi_{x_-}=\left(\frac{\epsilon B_0^-}{B_0^+}+\varphi_{0,x_-}\right)\psi,
\ee
where $\psi$ is a bodiless bosonic function of order $3$ of $x_-$.
\paragraph{}The first and second fundamental forms for this surface $\mathcal{S}$ are given by
\be
\hspace{-1.5cm}\ba{l}
I=d_+d_-\left[e^{\varphi_0-2\epsilon x_+}\left(1+\varphi_1\theta^+\theta^-\right)\psi\right],\\
I\hspace{-0.1cm}I=d_+^2\left[\psi^2e^{-\epsilon x_+}\left(B_0^++B_1^+\theta^+\theta^-\right)\right]+d_-^2\left[\psi^2e^{-\epsilon x_+}\left(B_0^-+B_1^-\theta^+\theta^-\right)\right]\\
\hspace{5cm}+d_+d_-\left[\psi e^{\varphi_0-\epsilon x_+}\left(h_0+\left(h_0\varphi_1+h_1\right)\theta^+\theta^-\right)\right].
\ea
\ee
Consequently, the Gaussian curvature and mean curvature are not constant. The Gaussian curvature is given by
\be
\hspace{-2.5cm}\mathcal{K}=e^{2\epsilon x_+}\left[\frac{\psi^2 \left(B_0^+B_0^-+\left(B_0^+B_1^-+B_0^-B_1^+\right)\theta^+\theta^-\right)}{e^{2\varphi_0}\left(1+2\varphi_1\theta^+\theta^-\right)}+\left(h_0^2+2h_0h_1\theta^+\theta^-\right)\right].
\ee
The umbilic points of the surface $\mathcal{S}$ occur when $\psi^2B_0^+B_0^-=0$ and 
\be
\psi^2(B_0^+B_1^-+B_0^-B_1^+)=0.
\ee

\paragraph{}\textbf{3.} For the subalgebra $L_{26}''=\lbrace K_1+(a-\frac{1}{2})K_0+\frac{1}{2}C_0,~a\neq\frac{1}{2}\rbrace$ we obtain the following parametrization of the orbit of the group
\be
\ba{ll}
H=(x_+)^{(a-1)/2}h(x_-,\eta,\theta^-),&\\
Q^+=(x_+)^{-(a+2)/2}q^+(x_-,\eta,\theta^-),& S^+=s^+(x_-,\eta,\theta^-),\\
Q^-=(x_+)^{-a/2}q^-(x_-,\eta,\theta^-),& S^-=(x_+)^{-1/2}s^-(x_-,\eta,\theta^-),\\
R^+=(x_+)^{-1/2}r^+(x_-,\eta,\theta^-),& T^+=(x_+)^{1/2}t^+(x_-,\eta,\theta^-),\\
R^-=(x_+)^{-1}r^-(x_-,\eta,\theta^-),& T^-=t^-(x_-,\eta,\theta^-),\\
e^\phi=(x_+)^{-a}\varphi(x_-,\eta,\theta^-),& f=(x_+)^{1/2}\psi(x_-),
\ea
\ee
where the bosonic symmetry variable is $x_-$ and the fermionic symmetry variables are $\eta=(x_+)^{-1/2}\theta^+$ and $\theta^-$. A corresponding invariant solution of the SUSY GC equations (\ref{GC}) takes the form
\be
\hspace{-2.5cm}\ba{l}
H=2iB(x_+)^{(a-2)/2}(\rho)_{x_-}\theta^+\theta^-,\\
Q^+=BA(x_-)(x_+)^{-(a+2)/2}\left[1+(x_+)^{-1/2}\theta^+\theta^-G(x_-)\right]\rho(x_-),\\
Q^-=\frac{2B}{a}(x_+)^{-a/2}\left[1+(x_+)^{-1/2}\theta^+\theta^-G(x_-)\right],\\
R^+=(x_+)^{-1/2}l_1\underline{R}_0^+,\hspace{0.3cm} R^-=(x_+)^{-1}l_2\underline{R}_0^-,\hspace{0.3cm} S^+=T^-=\underline{T}_0^-,\hspace{0.3cm} S^-=T^+=0,\\
e^\phi=A(x_-)(x_+)^{-a}(1+(x_+)^{-1/2}\theta^+\theta^-G(x_-)),\qquad f=(x_+)^{1/2}\psi(x_-),
\ea\label{quan1}
\ee
where $B=\underline{l}_0\underline{R}_0^+\underline{R}_0^-\underline{T}_0^-$ and $l_1,l_2,l_3$ are bosonic constants, while $\underline{l}_0$,  $\underline{R}_0^\pm$ and $\underline{T}_0^-$ are fermionic constants.
Here, $A$, $G$, $\rho$ and $\psi$ are arbitrary bosonic functions of the symmetry variable $x_-$. However, the function $A$ contains a part in $\Lambda_{body}$ but $\psi$ is a bodiless function.

The corresponding first and second fundamental forms for the surface $\mathcal{S}$ given by (\ref{quan1}) are
\be
I=\psi d_+d_-\left[A(x_+)^{-(2a+1)/2}\left(1+(x_+)^{-1/2}\theta^+\theta^-G\right)\right],
\ee
and
\be
\hspace{-2cm}\ba{l}
I\hspace{-0.1cm}I=(d_+)^2\left[AB(x_+)^{-(a+2)/2}\rho\left(1+(x_+)^{-1/2}\theta^+\theta^-G\right)\right]\\
\hspace{0.7cm}+2id_+d_-\left[AB(x_+)^{-1}\theta^+\theta^-\rho'\right]+(d_-)^2\left[\frac{2B}{a}(x_+)^{-a/2}\left(1+(x_+)^{-1/2}\theta^+\theta^-G\right)\right].
\ea
\ee
Consequently, the Gaussian curvature $\mathcal{K}$ and the mean curvature $H$ of the associated surface $\mathcal{S}$ are not constant. The Gaussian curvature is given by
\be
\mathcal{K}=\frac{8B}{aA}(x_+)^{a-1}\rho\left(1+(x_+)^{-1/2}\theta^+\theta^-G\right).
\ee
Since $H^2=0$, it follows that the surface $\mathcal{S}$ admits umbilic points along the curve defined by $\mathcal{K}=0$, which lies on the surface $\mathcal{S}$. The subalgebra of the Lie algebra for the classical GC equation (\ref{cCeq}) analogous to subalgebra $L_{26}''$ is $L_{1,2}'=\lbrace e_3+ae_0\rbrace$. The corresponding invariant solution is given by
\be
\ba{ll}
H(z,\bar{z})=l_0e^{-a(z+\bar{z})},\qquad Q(z,\bar{z})=k_0e^{a(z+\bar{z})},\\
U(z,\bar{z})=\frac{-2k_0}{l_0}e^{2a(z+\bar{z})},\qquad \bar{Q}(z,\bar{z})=k_0e^{a(z+\bar{z})},\qquad k_0,l_0\in\mathbb{R}.
\ea\label{classi}
\ee
In contrast to the SUSY case (\ref{quan1}), the Gaussian curvature $\mathcal{K}$ vanishes for the classical solution (\ref{classi}) associated with the subalgebra $L_{1,2}'$. In both cases however, the mean curvature $H$ is non-zero.

\section{Conclusions}\setcounter{equation}{0}
The objective of this paper was to construct a supersymmetric extension of the Gauss-Weingarten equations (\ref{GW}) and the Gauss-Codazzi equations (\ref{GC}) through a superspace and superfield formalism. The analysis included conformally parametrized surfaces immersed in a Grassmann superspace $\mathbb{R}^{(1,1\vert2)}$. This analysis allowed us to determine a Lie superalgebra of infinitesimal symmetries which generate Lie point symmetries of the SUSY GC equations (\ref{GC}). In addition, we also computed the Lie symmetry algebra of the classical GC equations (\ref{cCeq}) and classified the one-dimensional subalgebras of its largest finite-dimensional subalgebra into conjugacy classes. Comparing the symmetries of the SUSY GC equations (\ref{GC}) with those of the classical GC equations (\ref{cCeq}), we observe an additional dilation in the SUSY case. More specifically, $K_1$ and $K_2$ in (\ref{gen}) are supersymmetrized versions of the fields $e_3$ and $e_4$ in (\ref{ei}) respectively. The generators $K_0$ and $C_0$ in (\ref{gen}) play the role of center in the SUSY case in the same way that $e_0$ does in the classical case and we did not find any Virasoro algebras in the SUSY case. Next, we performed a group-theoretical analysis in order to classify all of the one-dimensional subalgebras of the obtained superalgebra (\ref{cla}) into conjugacy classes. Through the use of a generalized version of the symmetry reduction method we demonstrated for three subalgebras in (\ref{Li}) how to find exact invariant solutions of the SUSY model. A systematic use of the structure of the invariance supergroup of the SUSY GC equations (\ref{GC}) allowed us to generate (bosonic and/or fermionic) symmetry variables. For certain subalgebras, the invariants had a non-standard structure and therefore did not lead to invariant solutions. This phenomenon of non-standard invariants has also been observed in the analysis of symmetries of SUSY hydrodynamic-type equations. The SRM enabled us to reduce, after some transformations, the basic system of PDEs to many possible reduced PDE systems. We also explored certain geometrical properties of invariant solutions of the SUSY GC equations (\ref{GC}).

\paragraph{}This research could be extended in several other directions. One possibility would be to compute an exhaustive list of all symmetries of the SUSY GC equations and compare them to the classical case and also to apply the above SUSY extension methods to the SUSY GC system in higher dimensions. Due to the complexity of the computations involved, this would require the development of a computer algebra Lie symmetry package capable of handling odd and even Grassmann variables. To the best of our knowledge such a package does not presently exist. Conservation laws are well-established for the construction of the classical parametrized surfaces based on the generalized Weierstrass-Enneper formula for immersion \cite{Konopelchenko}, but it has been observed that, for the GC equations, such conservation laws are broken in their corresponding SUSY extensions. The problem of determining which quantities $Q^\pm$, $R^\pm$, $S^\pm$, $T^\pm$, $H$, $\mathcal{K}$, $\phi$, $f$ are conserved by the SUSY model still remains an open question for the GC equations. Another open problem is to determine whether all integrable SUSY systems possess non-standard invariants. These topics will be investigated in our future work.

\section*{Acknowledgements}
AMG's work was supported by a research grant from NSERC of Canada. SB acknowledges a doctoral fellowship provided by the FQRNT of the Gouvernement du Qu\'ebec. AJH wishes to acknowledge and thank the Mathematical Physics Laboratory of the Centre de Recherches Math\'ematiques for the opportunity to contribute to this research.

\end{document}